\documentclass[aps,pre,twocolumn,floatfix,10pt]{revtex4}
\usepackage{graphicx}
\usepackage{dcolumn}
\usepackage{bm}
\usepackage{latexsym}
\usepackage{amsmath}
\usepackage{amssymb}
\usepackage{color}
\usepackage[normalem]{ulem}
\usepackage{multirow}
\begin{document}
\title{{Level crossing statistics in a biologically motivated model of  \\  a long dynamic protrusion: passage times, random and  extreme excursions  }}
\author{Swayamshree Patra} 
\affiliation{Department of Physics, Indian
  Institute of Technology Kanpur, 208016, India} 
\author{Debashish Chowdhury{\footnote{Corresponding author: E-mail: debch@iitk.ac.in}}}
\affiliation{Department of Physics, Indian Institute of Technology
  Kanpur, 208016, India}
  \begin{abstract}
Long cell protrusions, which are effectively one-dimensional, are highly dynamic subcellular structures. Length of many such protrusions keep fluctuating about the mean value even in the the steady state. We develop here a stochastic model motivated by length fluctuations of a type of appendage of an eukaryotic cell called flagellum (also called cilium). Exploiting the techniques developed for the calculation of level-crossing statistics of random excursions of stochastic process, we have derived analytical expressions of passage times for hitting various thresholds, sojourn times of random excursions beyond the threshold and the extreme lengths attained during the lifetime of these model flagella. We identify different parameter regimes of this model flagellum that mimic those of the wildtype and mutants of a well known flagellated cell. By analysing our model in these different parameter regimes, we demonstrate how mutation can alter the level-crossing statistics even when the steady state length remains unaffected by the same mutation.  Comparison of the theoretically predicted level crossing statistics, in addition to mean and variance of the length, in the steady state with the corresponding experimental data can be used in near future as stringent tests for the validity of the models of flagellar length control. The experimental data required for this purpose, though never reported till now, can be collected, in principle, using a method developed very recently for flagellar length fluctuations.
\end{abstract}
  \maketitle
  
\section{Introduction}
The level-crossing phenomena \cite{masoliverbook} are characterized by the statistical distributions of varieties of quantities like passage times \cite{rednerbook}, exit and escape times \cite{masoliver14}, sojourn or residence times, extreme values, etc. \cite{braininabook}. These phenomena occur is many systems that collectively span several orders of magnitude of length scales and time scales. Therefore, theoretical methods of stochastic processes developed for calculation of the statistical characteristics of level-crossing phenomena have found applications in many physical, biological, environmental sciences as well as in engineering \cite{nordin70,ghusinga17,zhang16,polizzi16,metzler14,masoliver14,iyerbiswas16,guillet19,greulich18,malakar18,syski92,stratonovic81,braininabook,zacks17,
bel20,thorneywork19,dhar19,besga20,evans14,ghosh18}. 
The underlying physical kinetics that give rise to the fluctuating level crossing processes can be distinct in different systems. Motivated by an ubiquitous level-crossing phenomenon in molecular cell biology, we develop here a minimal model capturing some of the key elements of the underlying physical processes in this phenomenon. 

Our study is motivated by long cell protrusions like eukaryotic flagellum and cilium which are essentially one-dimensional dynamic structures \cite{patra20,chu17, mukherji14, mohapatra16, marshall01, amiri20,yuan17,herrmann16,pollard03,borisy16,piao09}. For proper biological function, size matters at all levels of biological organization \cite{haldane25,bonnerbook,hamant20,marshall15a,marshall15b,marshall16,
rafelski08} and in case of long cell protrusions, length is the dominant characteristic of their size. There are  strong experimental evidences \cite{ludington15} that suggest that the individual cells actively control the length  of these structures. It has also been established experimentally \cite{marshall01} that the length of dynamic protrusions like flagellum is maintained in the steady state at a constant mean value $\ell_\text{ss}$ by the balanced rates of the corresponding probabilistic assembly and disassembly. However, instantaneous length $\ell(t)$ of these dynamic structures fluctuates about the mean length due to the removal of the structural proteins and incorporation of fresh ones at the tip \cite{mohapatra16, albus13, folz19, ludington15, wordeman09, melbinger12, rank18, mohapatra15, varga09, orly15, patra20, prost07, govindan08, kuan13,reese14,erlenkamper09,fai19,banerjee20,ma20,johann12,klein05,gov06}. Consequently, in the steady state, the length is expected to remain confined within a narrow zone bounded by $ \ell_\text{ss}  \pm \sigma$ with variance $\sigma^2$. But, ongoing fluctuations often drive the length out of these bounds.

Using our recently developed theoretical model of flagellar length control, earlier \cite{patra20} we studied the correlations between the length fluctuations of the two flagella of a biflagellate cell both at and away from the steady-state. Using essentially the same model, here  we study the statistics of various level crossing quantities for a single flagellum in the steady state. Specifically, we address questions like: (a) What type of stochastic process are the flagellar length fluctuations? (b) What is the distribution of the first passage times for the flagellar length to  hit a given threshold? (c) What is the distribution of the durations for which  the length lies beyond these thresholds? (d) What are the maximum or minimum lengths that a flagellum can attain during its lifetime? Although the general expressions for some of these statistical quantities are known in the theory of stochastic processes, we report here the corresponding results specifically for our model of flagellar length control where the rate of the probabilistic assembly is length-dependent.

To our knowledge, no experimental data on level crossing statistics of temporally fluctuating flagellar lengths have been reported in the literature. Using the relevant  model parameters we characterize different aspects of flagellar length control in our model. Assigning a particular set of values to these parameters we could reproduce quantitative experimental data for the time-dependence of the average length of flagella in wildtype {\it Chlamydomonas} cells. Values of these parameters away from the corresponding wildtype values  are interpreted here to  mimic different types of mutants. We argue theoretically how level-crossing statistics would get affected in different type of mutants. We hope testing the validity of our theoretical predictions on the differences of level-crossing statistics in wildtype and mutant flagella may be possible in near future since experimental investigation of flagellar length fluctuations has already begun very recently \cite{bauer20}. 

One relevant question is: why are the level-crossing statistics important? There are potential utilities of these quantities in establishing the mechanism of flagellar length control. At present there are few distinct models, based on alternative scenarios, all of which can account for the experimentally observed time-dependence of the mean flagellar length. In case some distinct models predict the same mean steady-state length but different level crossing statistics in the steady state, it will be possible to rule out at least one (or more) of these, thereby narrowing the choice of the valid models, by comparing the theoretically predicted level crossing statistics with the corresponding experimental data. In other words, level-crossing statistics will impose additional stringent tests for the validity of the models of flagellar length control. Our work here is a step in that direction in the sense that we report the level-crossing statistics for a simple model of flagellar length control. We hope, in near future, these quantities will be calculated for other models of flagellar length control and will also be measured experimentally.

Temporal  length fluctuations of flagellum and cilium may have important implications in their biological functions\cite{mcgrath17,ortega19,brown10,snell04,mogilner96,mogilner03}. For example, if the fluctuating length of a  sensory cilium falls below a certain threshold it may not be able to pick up molecular signals floating beyond that threshold \cite{ferreira19}. When such a situation arises, the shortened cilium becomes silent; the consequent `{\it fading}' of chemical signals received by the cell would be reminiscent of the phenomenon of fading of electromagnetic signals that has been studied extensively \cite{braininabook,kratz06} since the pioneering work of Rice \cite{rice44,rice45}. The silent cilium regains its sensory capability only after its length grows above that threshold because of fluctuations. Similarly, growing beyond an upper threshold may cause swimming anomalies \cite{khona13} or lead to excessive exposure to the surrounding environment. This will add to the metabolic cost and disturb the energy budget of the cell \cite{milo16}. Therefore, understanding the steady state length fluctuations of protrusions belonging to both wild type cells and mutants in  a qualitative and quantitative manner from the perspective of level crossing statistics can be beneficial.

\begin{figure*}
\includegraphics[width=0.95\textwidth]{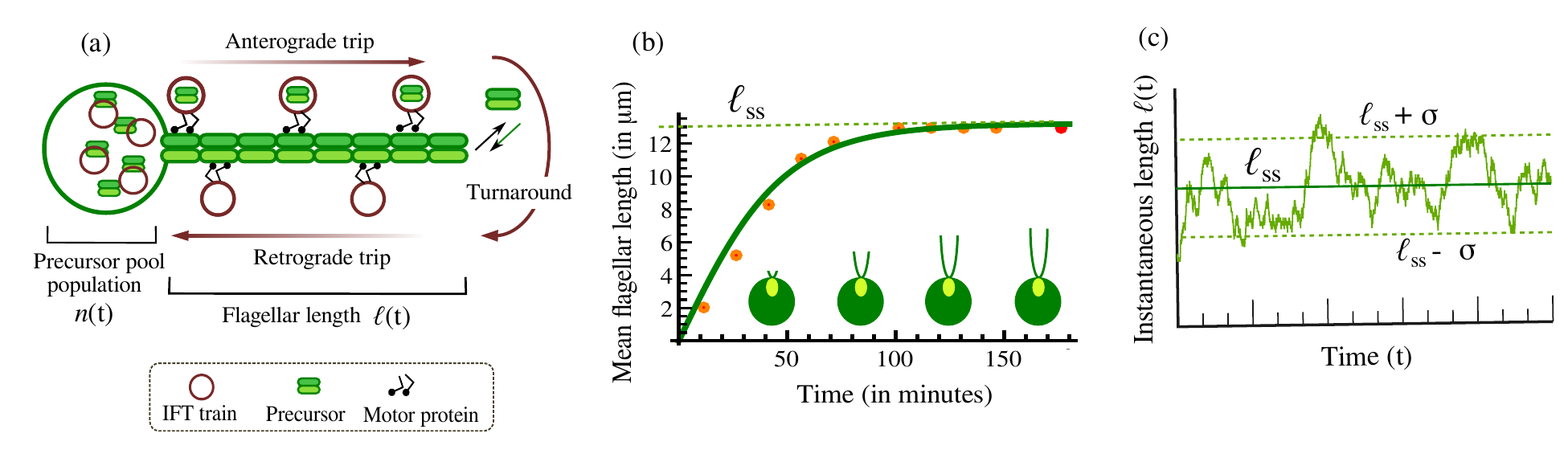}
\caption{ {\bf Evolution of flagellar length:} (a) Model for flagellar  assembly. Flagellum is denoted by a pair of parallel lattice chains and each lattice unit (or dimer) is the precursor. The precursors are synthesized and degraded in the pool. Fresh precursors are  transported by IFT trains  (pulled by motors) for incorporating them at the growing tip and discarded ones are brought back to the pool. (b) Mean flagellar length $\langle \ell(t) \rangle$ evolving with time obtained by solving the rate equation (\ref{flagella-rate-eq}). (Solid line corresponds to theory and dots correspond the experimental data from ref.\cite{ishikawa17}. (c) Instantaneous flagellar length $\ell(t)$ fluctuating about the steady state mean value $\ell_\text{ss}$ obtained by simulating the model with Monte Carlo methods. Two natural bounds for protrusions of controlled length are $\ell_\text{ss} \pm \sigma$ denoted by horizontal lines.  Parameters used for plot in (b-c) are $\rho=0.08$, $J=0.0736$, $v=0.92$, $\Omega_e=0.65$, $\Gamma_r=3.0\times 10^{-4}$, $k=1.0\times10^{-3}$, $\omega^+=4.5 \times 10^{-4}$, $\omega^-=4.5 \times 10^{-6}$, $n_\text{max}=500$, $\delta t=3.6\times 10^{-4}$ s. }

\label{fig_model}
\end{figure*}

\section{Temporal length fluctuations of eukaryotic flagellum and cilium }
\label{sec-flagellum}

\subsection{Model for flagellar length control}
Eukaryotic flagellum is a membrane bound organelle that projects out of the cell body. Nine microtubule (MT) doublets arranged in a cylindrical fashion with the two singlets at the center form the primary structural scaffold of the flagellum. The flagellum elongates by adding structural proteins (now onwards, loosely, referred to as ``precursor'') to its growing tip but because of the turnover of tubulins at the flagellar tip, the flagellum shortens by removal of  precursors \cite{marshall01,lechtreck17,ludington15,ishikawa17}. The flagellum exchanges its precursor with a precursor pool situated at the flagellar base which  contains the machineries for synthesis and degradation of the precursors \cite{lechtreck17}. An active transport mechanism known as intraflagellar transport (IFT) is responsible for facilitating this exchange of precursors\cite{engel09,kozminski93,kozminski12,rosenbaum02}. IFT trains consist of IFT proteins which are arranged in linear arrays just like the bogies of a train. The IFT trains loaded with precursors are driven towards the flagellar tip by the kinesin motors walking on one of the MTs of the doublet (B-MT). At the tip, the trains get `remodelled', kinesins are `disengaged' and the dynein motors get `activated'. Thereafter, the IFT trains loaded with discarded material pulled back to the base by the dynein motors walking on the other MT of the doublet (i.e, 
A-MT)\cite{stepanek16}.

Our model for flagellar assembly is shown in Fig.\ref{fig_model}(a). The flagellum is represented as a pair of antiparallel lattices, where each lattice represents a MT of the doublet. IFT trains pulled by the motors are represented by self driven particles which obey exclusion principle and jump  to the neighboring sites in a stochastic manner with a certain hopping rate.  The flux of the IFT particles inside the flagellum is $J$, their number density on the lattice is $\rho$ and the effective velocity with which they move is $v$. Each IFT particle can be either empty or loaded with a single precursor (lattice unit) which can elongate the model flagellum (pair of lattices) at the tip by a single unit. The cell senses the flagellar length with a {\it time of flight} mechanism. Suppose, the timer molecule, which is an integral part of the IFT trains, enters the flagellum in a particular chemical (or conformational) state. The probability of finding the timer in the same state after time $t$ is given by $e^{-k t}$ where $k$ is rate of flipping of the state of the timer.  So, if the mean flagellar length is $\langle \ell (t) \rangle$ at time $t$, the probability of the timer returning to the base (after the roundtrip inside flagellum) without flipping state is $e^{-k t_{tof}}$ where $t_{tof}$ is the time of flight and is given by $t_{tof}=2\langle \ell (t) \rangle/v$. Hence, if the timer returns without flipping its state with probability  $e^{-k t_{tof}}$, the IFT train with a precursor is dispatched otherwise empty IFT train enters the flagellum. Whether the IFT train is loaded or not with a precursor also depends on the amount of precursor in the pool. If the current average  population of pool is $\langle n (t) \rangle$ and its maximum capacity  is $n_\text{max}$, then the probability of loading of precursors onto an IFT train is proportional to $\langle n (t) \rangle/n_\text{max}$. Therefore, the flux of full trains reaching the flagellar tip is
\begin{equation}
J_{full}=\frac{ \langle n (t) \rangle} { n_\text{max}} ~ e^{-2k \langle \ell(t)\rangle /v} J~.
\end{equation}
The precursor loaded on the IFT train can elongate the flagellum with probability $\Omega_e$. So the overall rate of assembly is given by
\begin{equation}
\frac{\langle n(t) \rangle}{n_\text{max}} J \Omega_e ~ e^{-2k \langle \ell(t)  \rangle /v}.
\label{eff-ass-rate}
\end{equation} 
Due to ongoing turnover, if both the sites at the tip are not occupied by any IFT trains, the dimer (i.e the precursor) dissociates with rate $\Gamma_r$. Therefore, the overall disassembly rate is given by
\begin{equation}
(1-\rho)^2 \Gamma_r~.
\label{eff-dis-rate}
\end{equation}
The pool synthesises and degrades  precursors in a population dependent fashion so that precursor population does not exceed $n_\text{max}$  with rate $\omega^+$ and $\omega^-$ respectively. The coupled set of equations which govern the evolution of mean flagellar length $\langle \ell (t) \rangle$ and pool population $\langle n (t) \rangle $ are 
\begin{eqnarray}
\frac{d\langle \ell (t) \rangle}{dt} &=& \frac{\langle n (t) \rangle}{n_\text{max}}J\Omega_e \exp \left(  -\frac{2k \langle \ell (t) \rangle }{v} \right) -(1-\rho)^2 \Gamma_r \nonumber \\
\frac{d \langle n (t) \rangle}{dt}&=&\omega^+ \left( 1-\frac{\langle n (t) \rangle}{n_\text{max}} \right) - \omega^- \langle n (t) \rangle~.
\label{flagella-rate-eq}
\end{eqnarray}
As done in our previous analysis, we solve for the evolution of mean length $\langle \ell (t) \rangle$ of the flagellum of wild type cell of {\it Chlamydomonas reinhardtii} using the set of coupled equation (\ref{flagella-rate-eq}) and multiply it with 8 nm for converting it into actual length; 8 nm being the length of a single MT dimer \cite{patra20} (see Fig.\ref{fig_model}(b)).  We chose this species because the flagellum of this particular species of green algae is most commonly used in experimental studies of length control  \cite{marshall01,ludington15,ishikawa17}. Now, we will use this model for exploring the statistical properties of steady state flagellar length fluctuations. 

\subsection{Mapping of flagellar length fluctuations onto Ornstein-Uhlenbeck (OU)  process}
The growth and shortening of the dynamic flagellum by the addition and removal of precursors at the distal tip are intrinsically stochastic \cite{marshall01}. Further stochasticities arise from the noisy synthesis and degradation of the precursors in the cell body \cite{banerjee20}, stochastic loading of precursor  into the IFT trains \cite{wren13}, entry of IFT trains into the flagellum \cite{ludington13, bressloff18} and their transport \cite{stepanek16,patra18,bressloff06}. These stochasticities collectively result in the instantaneous flagellar length  fluctuating about the mean value as shown in Fig.\ref{fig_model}(c).

For the quantitative description of the elongation and shortening dynamics of the flagellum, we treat it as a stochastic process $L(t)$ where the stochastic kinetics of the length of the protrusion  are assumed to be Markovian. The master equation governing the stochastic elongation/shortening kinetics \cite{kampen10,gardiner09,gillespie13} of the protrusion is given by
\begin{eqnarray}
\frac{dP_L(\ell,t)}{dt}&=&r^+(\ell-1)P_L(\ell-1,t)+r^-(\ell+1)P_L(\ell+1,t)\nonumber\\&-&(r^+(\ell)+r^-(\ell))P_L(\ell,t)  
\label{master-eq-gp-mt}
\end{eqnarray} 
where $P_L(\ell,t)$ is the probability of  having a protrusion of length $L(t)=\ell$ at time $t$ and $\ell$ is a positive integer. From expression (\ref{eff-ass-rate}) and (\ref{eff-dis-rate}), the length dependent assembly $(r^-(\ell))$ and length independent disassembly rate $(r^-(\ell))$ are given by
\begin{equation}
r^+(\ell)=A e^{-C \ell}  ~~ ~{\rm and}~ ~~ r^-(\ell)=B
\label{flagella-rates}
\end{equation}
where
\begin{equation}
A=\frac{  n_\text{ss} }{ n_\text{max} }J\Omega_e, ~~ B=(1-\rho)^2 \Gamma_r ~~ \text{and} ~~ C=2k/v ~.
\label{eq-ABCflag}
\end{equation}
The steady state mean length emerges when $r^+(\ell)=r^-(\ell)$ and  is given by
\begin{equation}
\ell_\text{ss}=\frac{v}{2k} \ell og \left[ \frac{  n_\text{ss} }{ n_\text{max} } \frac{J\Omega_e}{(1-\rho)^2 \Gamma_r} \right] = \frac{1}{C}\ell og \left[ \frac{A}{B} \right]~.
\label{lss}
\end{equation}
From the set of parameters used to for the plot in Fig.\ref{fig_model}(b) (mentioned in the caption of Fig.\ref{fig_model}) and using the expressions in eq.\ref{eq-ABCflag} we get
\begin{eqnarray}
A_\text{Wt}= 809.6 ~\text{min}^{-1} , B_\text{ Wt} = 28.21 ~\text{min}^{-1}~\nonumber\\
\text{and}~ C_\text{Wt} = 0.0021
\label{eq-wt}
\end{eqnarray}
where the subscript Wt indicates the value of parameters corresponding to flagella of the wild type cell. 

In the continuum limit, the time evolution of the probability density of the protrusion length is governed by the corresponding Fokker-Planck equation \cite{kampen10,gardiner09} 
{\small{
\begin{eqnarray}
\frac{\partial p(y,t)}{\partial t}=-\frac{\partial }{\partial y} \bigg{[} R_{-}(y) p(y,t)\bigg{]}+\frac{1}{2N}\frac{\partial^2 }{\partial y^2} \bigg{[} R_{+}(y) p(y,t)\bigg{]}
\label{fpe-eq-gp}
\end{eqnarray} }}
where the length is defined by a continuous variable $y$ and the drift  and the diffusion are given by $R_{-}(y)(=r^{+}(y) - r^{-}(y))$ and $R_{+}(y)(=r^{+}(y) - r^{-}(y)))$, respectively.

\begin{figure}
\includegraphics[width=0.38\textwidth]{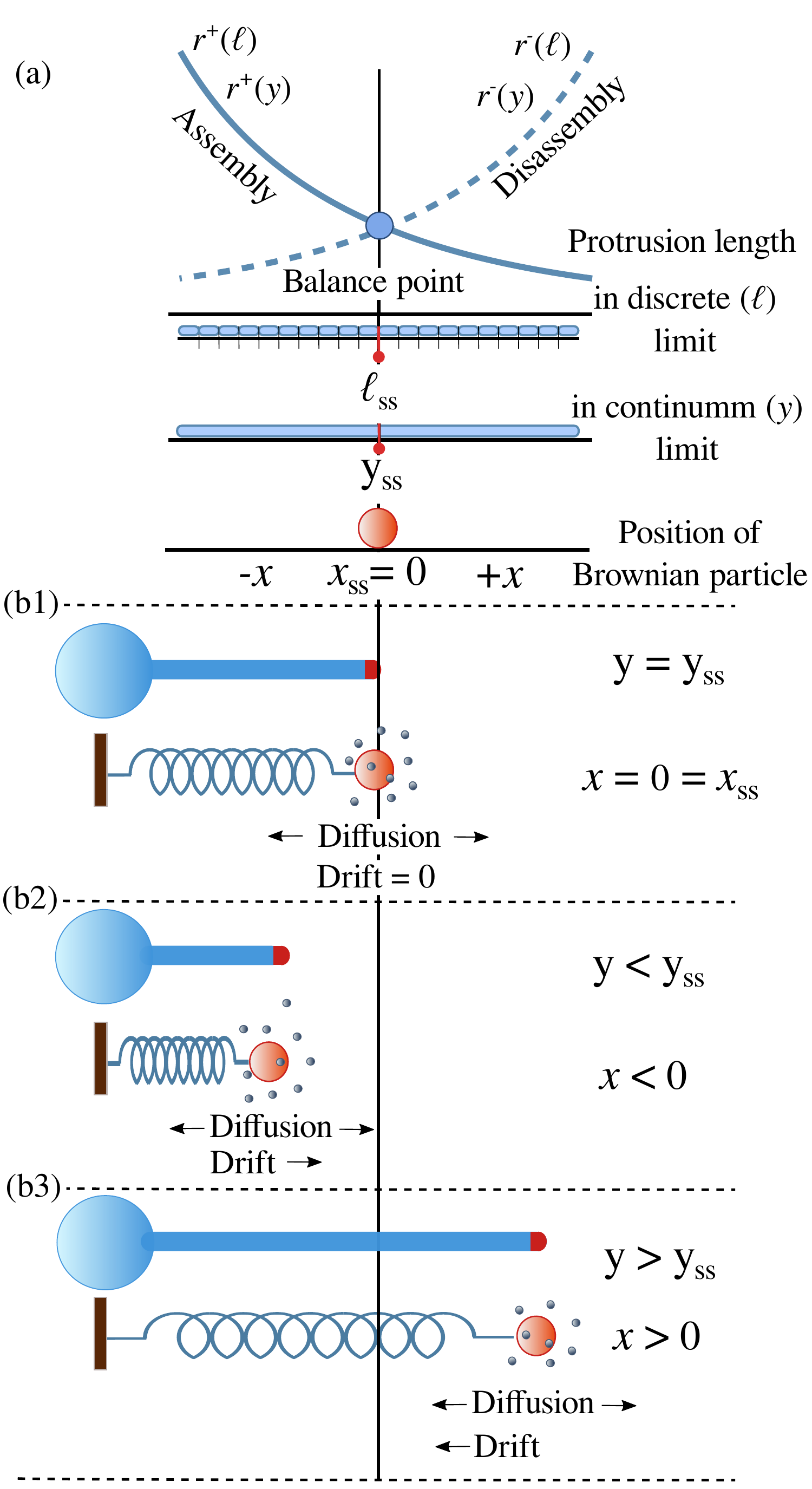}
\caption{ {\bf Mapping length fluctuations into {\it Ornstein-Uhlenbeck} process :} (a) Balance point emerges in the discrete and continumm limit wherever the assembly and disassembly rates intersect and the corresponding length is the steady state mean length of the protrusion ($\ell_\text{ss}$ in discrete limit and $y_\text{ss}$ in continumm limit). The position of the tip of the protrusion can be mapped into a hypothetical Brownian particle whose mean position is at $x_\text{ss}=0$. (b) Mapping the position of the tip of the protrusion with fluctuating length into the position of a Brownian particle which is attached to a spring. In both the cases diffusion drives the tip $y$ or the Brownian particle ($x$) away from the steady state position but the restoring drift tends to restore their position towards the steady value ($y_\text{ss}$ and $x_\text{ss}$ respectively).}
\label{mapping}
\end{figure}

As we are interested in studying the properties of steady state length fluctuations, we make a change of variable from $y$ to $x$  by defining $y-y_\text{ss}=x/\sqrt{N}$ where $x$ is a measure of the deviation of $y$ from its steady-state value $y_\text{ss}$ (see Fig.\ref{mapping}). The variable $x$ obeys the following Fokker-Planck equation
\begin{eqnarray}
\frac{\partial p(x,t)}{\partial t}=\frac{\partial }{\partial x} \bigg{[} \kappa  x \ p(x,t)\bigg{]}+\frac{D}{2}\frac{\partial^2 p(x,t)}{\partial x^2}
\label{FPE-OU-gp}
\end{eqnarray}
where 
\begin{equation}
\kappa= - R'_{-}(y_\text{ss})=BC= ~ \&  ~ D=R_{+}(y_\text{ss})=2B~.
\label{def_kappa_D}
\end{equation}
(See appendix A for the main steps of the derivation.)

\begin{table*}
\caption{Statistics of level crossing quantities for flagellar length fluctuations}
\label{OU_Table}
\begin{tabular}{ |c|c| }
\hline
\hline 
\textbf{Formula} & \textbf{Timescales}\\
\textbf{No.} & \\
\hline
\hline
1.1 & {Mean escape time}\\
\hline
& \\
& $\mathcal{T}_{\mathcal{E}}(x_U,x_L|\underline{x_0,t_0})=$ \\
& ${\bigg{[}D \left(\text{erfi}\left(\frac{\sqrt{\kappa } x_L}{\sqrt{D}}\right)-\text{erfi}\left(\frac{\sqrt{\kappa } x_U}{\sqrt{D}}\right)\right)\bigg{]}}^{-1} \times
 \bigg{[} x_0^2 \left(\text{erfi}\left(\frac{\sqrt{\kappa } x_U}{\sqrt{D}}\right)-\text{erfi}\left(\frac{\sqrt{\kappa } x_L}{\sqrt{D}}\right)\right) \, _2F_2\left(1,1;\frac{3}{2},2;\frac{\kappa  x_0^2}{D}\right)$ \\ 
 & $+x_L^2 \left(\text{erfi}\left(\frac{\sqrt{\kappa } x_0}{\sqrt{D}}\right)-\text{erfi}\left(\frac{\sqrt{\kappa } x_U}{\sqrt{D}}\right)\right) \, _2F_2\left(1,1;\frac{3}{2},2;\frac{\kappa  x_L^2}{D}\right)+x_U^2 \left(\text{erfi}\left(\frac{\sqrt{\kappa } x_L}{\sqrt{D}}\right)-\text{erfi}\left(\frac{\sqrt{\kappa } x_0}{\sqrt{D}}\right)\right) \, _2F_2\left(1,1;\frac{3}{2},2;\frac{\kappa  x_U^2}{D}\right)\bigg{]} $\\
 & where $x_\text{0}=\ell_0-\ell_\text{ss}$, $x_\text{U}=\ell_U-\ell_0$ and $x_\text{L}=\ell_0-\ell_L$\\
 & \\
\hline
1.2 & {Mean upcrossing time}\\
\hline
& \\
& $ \mathcal{T}_{\mathcal{HU}}(x_\text{th}|\underline{x_0,t_0})=\frac{1}{D} \bigg{[}{  x_{\text{th}}^2 \, _2F_2\left(1,1;\frac{3}{2},2;\frac{\kappa  x_{\text{th}}^2}{D}\right)- x_0^2 \, _2F_2\left(1,1;\frac{3}{2},2;\frac{\kappa  x_0^2}{{D}}\right)}\bigg{]} $  \\
& where $x_\text{th}=|\ell_\text{th}-\ell_\text{ss}|$ and $x_\text{0}=\ell_\text{0}-\ell_\text{ss}$. \\ 
& \\
\hline
1.3 & {Mean downcrossing time}\\
\hline
& \\
& $\mathcal{T}_{\mathcal{HD}}(x_\text{th}|\underline{x_0,t_0})
=\frac{1}{{2 \kappa }} \bigg{[} {\pi  \text{erfi}\left(\frac{\sqrt{\kappa } x_0}{\sqrt{D}}\right)-\pi  \text{erfi}\left(\frac{\sqrt{\kappa } x_{\text{th}}}{\sqrt{D}}\right)} \bigg{]} -\frac{1}{{D}}\bigg{[}{x_0^2 \, _2F_2\left(1,1;\frac{3}{2},2;\frac{\kappa  x_0^2}{D}\right)-x_{\text{th}}^2 \, _2F_2\left(1,1;\frac{3}{2},2;\frac{\kappa  x_{\text{th}}^2}{D}\right)} \bigg{]}$ \\
& where $x_\text{th}=|\ell_\text{th}-\ell_\text{ss}|$ and $x_\text{0}=\ell_\text{0}-\ell_\text{ss}$. \\
& \\
\hline
1.4 & {Mean number of sojourns  $n(\kappa \tau^+)$ of duration longer than $\tau^+$ per unit time }\\
\hline
& \\
& $n(\tau^+)=\frac{1}{2} \sqrt{\frac{\kappa}{D}} \exp \left( -\frac{\kappa x^2}{D} \right)    \bigg{[ } \sqrt{\frac{2D}{{\pi \tau^+ }}}~{e^{-(\kappa x_\text{th})^2 \tau^+/(2D)}}  - (\kappa x_\text{th}) ~ \text{erfc} \bigg{(} \kappa x_\text{th} \sqrt{\frac{\tau^+}{2D}} \bigg{)} \bigg{]}$ \\
& where $x_\text{th}=|\ell_\text{th}-\ell_\text{ss}|$\\
& \\
\hline
\hline
\textbf{Formula} & \textbf{Lengthscales}\\
\textbf{No.} & \\
\hline
\hline
2.1 & Maximum and minimum length\\
\hline
& \\
& $\langle {x}_\text{max} (t) |0 \rangle \simeq \sqrt{{\pi} Dt}$ ~~~and~~~ $\langle {x}_\text{min} (t) |0 \rangle=-\langle {x}_\text{max} (t) |0 \rangle=- \sqrt{\pi D t}$\\
& where $x_\text{max}(t)=\ell_\text{max}(t)-\ell_\text{ss}$ and $x_\text{min}(t)=\ell_\text{ss}-\ell_\text{min}(t)$\\
& \\
\hline
2.2 & Range scanned\\
\hline
& \\
& $\langle x_{range}(t)|0 \rangle=\langle {x}_{max} (t) |0 \rangle -\langle {x}_{min} (t) |0 \rangle =2\sqrt{ \pi Dt }$\\
& where ${x}_{range}=\ell_\text{range}=\ell_\text{max}-\ell_\text{min}$\\
& \\
\hline
\end{tabular}
\end{table*}

Thus, rescaling the flagellar length ($y-y_\text{ss}=x/\sqrt{N}$) we have mapped the length kinetics onto an Ornstein Uhlenbeck (OU) process (eq. (\ref{FPE-OU-gp})). This mapping allows us to describe the stochastic movement of the flagellar tip (variable $y$) about its mean position $y_\text{ss}$ due to the probabilistic elongation and shortening of the flagellum in terms of the one-dimensional kinetics of an overdamped Brownian particle, whose position (variable $x$) coincides with the flagellar tip (see Fig.\ref{mapping}(a-b)) and fluctuates about the steady state mean position $x_\text{ss}=0$. The Brownian particle is subjected to a spring force, $\kappa$ being the spring constant. This  mapping is consistent due to the well known fact that the drift velocity of an overdamped particle is proportional to the force (see Fig.\ref{mapping}(b1-b3)). This spring force and the restoring drift velocity $R_{-}(y)$ are equivalent, but alternative ways of describing the process that  constrains length fluctuations observed experimentally by Bauer et al.\cite{bauer20}. The `confining effect' of the drift velocity $R_{-}(y)$ mentioned in the context of flagellum and visualized in Fig.\ref{fig_model}(c), is equivalent to the `restoring effect' of the spring force $\kappa x$ (see Fig.\ref{mapping}(b1-b3)). 

A strong experimental support for mapping the length fluctuations into OU process is provided by Bauer et al \cite{bauer20}.  By monitoring the flagellar length in real time, they concluded that the motion of the  tip due to growing and shortening about the mean length resembles that of a Brownian particle which diffuses  but is also attached to a spring  which tend to restore its position to the mean value. Using our model for flagellar length control, we have shown how the two essential parameters of OU process $\kappa$ and $D$ depends on the microscopic properties of flagellum (see equation (\ref{def_kappa_D}) and (\ref{eq-ABCflag})).  $\kappa$ depends on  $B$ and $C$. $B$ depends on the number density of the trains on the track $\rho$ and the rate of the dissociation of the subunit at the tip of the protrusion $\Gamma_r$ whereas $C$ depends on the the velocity of travel $v$ of a timer
(which is identical to that of an IFT train) and the rate of flipping of the timer $k$.  On the other hand,  $D$ depends solely on $B$ which is the effective disassembly rate. Expression (\ref{lss}) for the $\ell_\text{ss}$ indicates that $A, B$ and $C$ are three key parameters of the model that determine the length of the flagellum in the NESS of the system whereas the equation (\ref{def_kappa_D}) indicate that only the parameters $B$ and $C$ determine the properties of steady state length fluctuations about the mean value $\ell_\text{ss}$. 

\begin{figure*}
\includegraphics[width=0.95\textwidth]{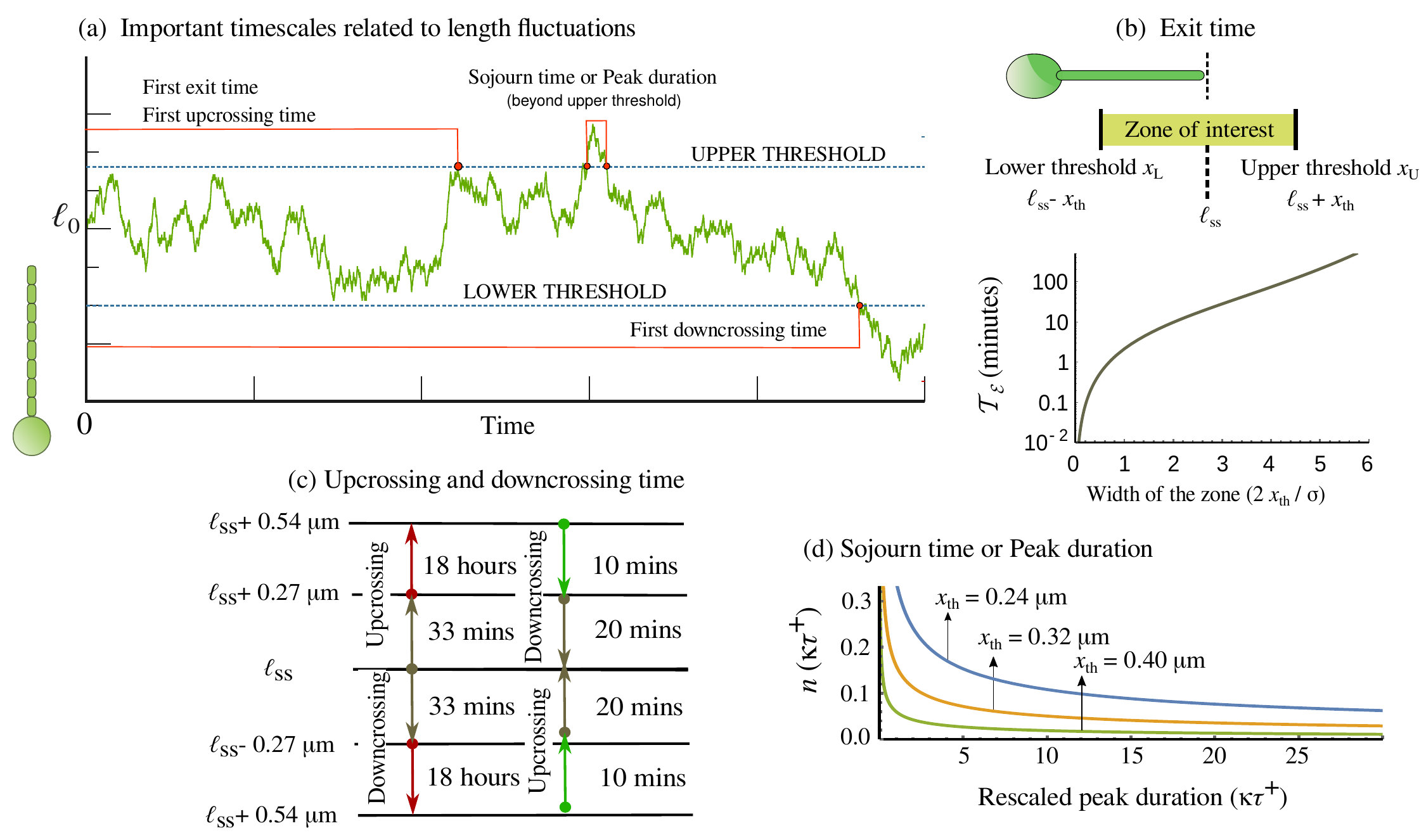}
\caption{Timescales: (a) A trajectory depicting the evolution of instantaneous flagellar length with time. Various passage times like first exit time, upcrossing and downcrossing time and sojourn time or peak duration of a peak are denoted. (b) Mean escape time $\mathcal{T}_\mathcal{E}$ as a function of width of  symmetric  zone. (c) Estimates of mean upcrossing time and downcrossing time for hitting special pair of thresholds. (d)  Number density of  peak $n(\kappa \tau^+)$ is plotted as function of the rescaled sojourn time or peak duration $\kappa \tau^+$  for three different thresholds (i) $x_{th}=0.24\mu$m (ii) $x_{th}=0.36\mu$m and (iii) $x_{th}=0.40\mu$m  for the flagellum of wild type cell.  }
\label{fig_time}
\end{figure*}

\section{Level crossing statistics for flagellar length fluctuations}

The direct benefit of mapping the protrusion length fluctuations onto the OU process is that now for the systematic analysis of length fluctuations, we can use the level crossing techniques developed for OU process where the Brownian particle is subjected to a linear restoring drift. The list of formulae for the statistics of various level crossing quantities which will be discussed in the context of flagellar length fluctuations is summarized in Table.\ref{OU_Table}.  For the convenience of the readers, we have presented the main steps in their derivation in the appendices (B-D). 
 
\subsection{Relevant timescales}

The four relevant timescales are the exit time, upcrossing time,  downcrossing time and sojourn time (see Fig.\ref{fig_time}(a)). The steps for the derivation of the respective expressions (1.1-1.4) of Table.\ref{OU_Table} that characterize their statistics are given in appendices B and C.

{\bf Exit time:} If initially, the flagellar length $\ell_0$ or the position of the tip  lies in a  zone bounded by an upper $\ell_U$ and lower $\ell_L$ threshold ($\ell_L \leq \ell_0 \leq \ell_U$), the time to escape (or exit) the  zone for the first time by crossing either of the two thresholds is called the {\it first exit time} or simply the exit time (see Fig.\ref{fig_time}(a)). For the calculation of mean exit time for the flagellar tip, we  define the ``zone of interest''  (see Fig.\ref{fig_time}(b));  the  two thresholds bounding the zone are placed symmetrically at $\ell_U=\ell_\text{ss}+x_\text{th}$ and $\ell_L=\ell_\text{ss}-x_\text{th}$. The width of the zone of interest can be varied by varying $|x_{th}|$. If the initial position of the  $\ell_0=\ell_\text{ss}$, using the formula (1.1) in table.\ref{OU_Table}, the expression for mean exit $\mathcal{T}_{\mathcal{E}}$ time is given by
{\small
\begin{eqnarray}
\mathcal{T}_{\mathcal{E}}(x_\text{th},x_\text{th}|\underline{\ell_{ss},0}) = \frac{ x_\text{th}^2 }{D}\, _2F_2\left(1,1;\frac{3}{2},2;{\frac{\kappa}{D}} x^2_\text{th} \right)~.
\label{met-eq-th1}
\end{eqnarray} }
Using the expansion \cite{abramowitz72}
\begin{equation}
 _2F_2\left(1,1;\frac{3}{2},2;z\right)=1+\frac{z}{3}+\mathcal{O}(z^3)
\end{equation}
in the  expression (1.1) it is easy to check that for narrow zones with  $x^2_\text{th}<\sigma$
\begin{equation}
\mathcal{T}_{\mathcal{E}}(x_\text{th},x_\text{th}|\underline{\ell_{ss},0}) \approx \frac{1}{D}(x_\text{th}^2)=\frac{x_\text{th}^2}{2B}=\frac{x_\text{th}^2}{2(1-\rho)^2{\Gamma_r}}
\label{eq-OnlyDdep1}
\end{equation} 
i.e the $\mathcal{T}_{\mathcal{E}} \propto x^2_\text{th}$
and for broader zones with  $x^2_\text{th}>\sigma$
\begin{equation}
\mathcal{T}_{\mathcal{E}}(x_\text{th},x_\text{th}|\underline{\ell_{ss},0})  \approx \frac{\kappa}{3 D^{2}}(x_\text{th}^4)=\frac{  C x_\text{th}^4 }{12 B}=\frac{k x_\text{th}^4}{6 v (1-\rho)^2 \Gamma_r}.
\label{eq-OnlyDdep2}
\end{equation} 
i.e, the $\mathcal{T}_{\mathcal{E}} \propto x^4_\text{th}$. Note the associated variance is given by
\begin{equation}
\sigma^2=\frac{D}{2 \kappa}
\end{equation}
We have plotted $\mathcal{T}_{\mathcal{E}}$ as a function of the width of safe zone $2x_\text{th}$ as indicated in Fig.\ref{fig_time}(b) in a semilog plot and as indicated by the expressions (\ref{eq-OnlyDdep1}) and (\ref{eq-OnlyDdep2}), it can be observed how the slope varies as the width of the region is increased.

To understand the physical origin of this observation, note that the equation (\ref{eq-OnlyDdep1}) has been obtained in the limit of small $x_\text{th}$ ($x_\text{th}/\sigma<1$). The deviation from the mean value and escape from narrow zone is caused by the random diffusion associated with the fluctuations. But for broader zones with larger values of $x_\text{th}$, significant restoring drift $\kappa x$ directed towards the mean arises due to active sensing of the flagellar length and this makes escape driven by the diffusion challenging as compared to the escape from a narrow zone. This is indicated both by the equations (\ref{eq-OnlyDdep1}) and (\ref{eq-OnlyDdep2}) and also by the curve that changes slope for values $2x_\text{th}/\sigma>1$ and $2x_\text{th}/\sigma<1$.

\begin{figure*}
\includegraphics[width=1.0\textwidth]{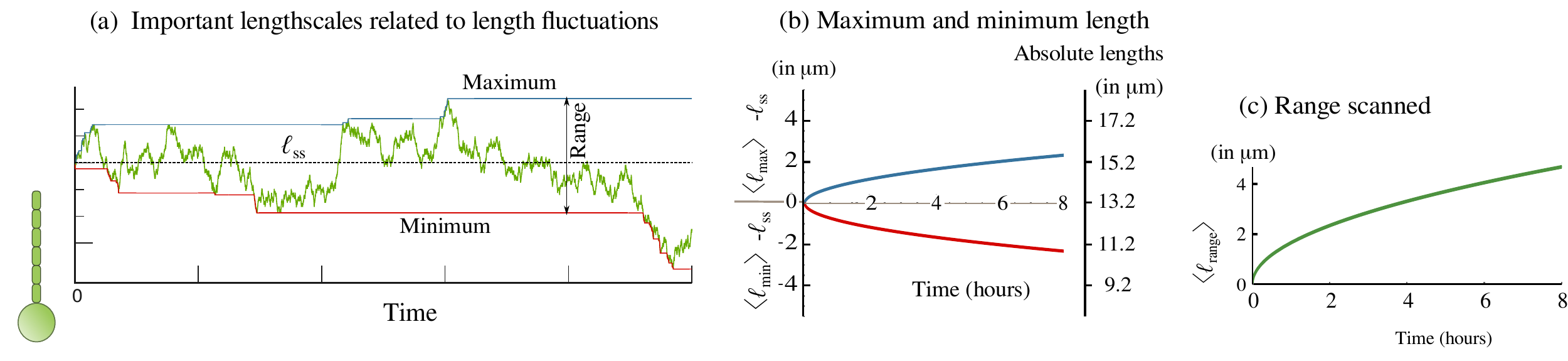}
\caption{Lengthscale: (a) The  average maximum $\ell_\text{max}(t)$ and minimum $\ell_\text{min}(t)$ lengths attained (b) and the average width of the range $\ell_\text{range}(t)$ scanned by the flagellum of  the wild type cell about their respective steady state mean length $\ell_\text{ss}$ as a function of time.  }
\label{fig_len}
\end{figure*}

{\bf Upcrossing and downcrossing times:} In case there is a single threshold $\ell_\text{th}$  of interest, there are two ways of hitting it  (see Fig.\ref{fig_time}(a)). If the initial length  $\ell_0$ lies below the threshold $\ell_\text{th}$ as shown in Fig.\ref{fig_time}(a), then the length must increase  to hit the threshold length $\ell_\text{th}$ and the first time it hits the threshold from below is known as the first upcrossing time $ \mathcal{T}_{\mathcal{HU}}(\ell_\text{th}|\underline{\ell_0,0})$. On the other hand, if $\ell_0$ lies above $\ell_\text{th}$ (Fig.\ref{fig_time}(a)), then length shortens  to hit the threshold $\ell_\text{th}$; the first time it hits the threshold $\ell_\text{th}$ from above is the first downcrossing time $ \mathcal{T}_{\mathcal{HD}}(\ell_\text{th}|\underline{\ell_0,0})$.

Naively, one might expect the mean time for  upcrossing the threshold at $\ell_\text{th}$ from $\ell_0 ~( < \ell_\text{th})$ to be identical to mean time for downcrossing the same threshold from $\ell_0 ~ (> \ell_\text{th})$ if the distance $|\ell_0-\ell_\text{th}|$ is same in both the cases. But, that is not true, as the expressions for mean upcrossing time and downcrossing time (see formula (1.2) and (1.3) of Table.\ref{OU_Table}) and detailed analysis of hitting times for the flagellar length indicate. Using the expressions (1.2) amd (1.3) of Table.\ref{OU_Table}, we estimate the mean hitting times for the pair of thresholds which are shown in Fig.\ref{fig_time}(c). The magnitude of mean upcrossing time from $\ell_\text{ss}$ to $\ell_\text{ss}+x_\text{th}$ is same as the magnitude of mean downcrossing time from $\ell_\text{ss}$ to $\ell_\text{ss}-x_\text{th}$. This symmetry is a consequence of the symmetry of the diffusion and restoring drift on both sides of $\ell_\text{ss}$. For the same reason the mean upcrossing time from $\ell_\text{ss}+x_\text{th}$ to $\ell_\text{ss}+2x_\text{th}$ is same as the magnitude of mean downcrossing time from
$\ell_\text{ss} - x_\text{th}$ to $\ell_\text{ss} - 2x_\text{th}$. Since the restoring drift (or, equivalently, spring force) increases with increasing distance from the mean position, the mean upcrossing time from $\ell_\text{ss}+x_\text{th}$ to $\ell_\text{ss}+2x_\text{th}$ is much longer than that from $\ell_\text{ss}$ to $\ell_\text{ss}+x_\text{th}$ whereas the mean downcrossing time from $\ell_\text{ss}+2x_\text{th}$ to $\ell_\text{ss}+x_\text{th}$ is much shorter than that from $\ell_\text{ss}+x_\text{th}$ to $\ell_\text{ss}$.

{\bf Sojourn time:} The sojourn times or peak duration above a certain threshold $\ell_\text{th}$ is denoted by $\tau^+$ (see Fig.\ref{fig_time}(a)). A class of important thresholds $\ell_\text{th} (\ell^2_\text{th} >> \sigma^2(=D/2\kappa ))$ are the ones which are rarely visited by the tip of the flagellum of fluctuating length. For such high thresholds, we estimate the mean number of sojourns  $n(\kappa \tau^+)$ of duration longer than $\tau^+$ per unit time by using the expression (1.4) in Table.\ref{OU_Table}. For the flagellum of wild type cell,  the mean number of sojourns  $n(\kappa \tau^+)$ of duration longer than $\tau^+$ per unit time is plotted as function of the rescaled time $\kappa \tau^+$ in Fig.\ref{fig_time}(c) for three different thresholds $x_\text{th}$. Irrespective of the threshold value $x_{th}$, $n(\tau^+)$ decays with increasing sojourn time $\tau^+$. Besides, for a given rescaled peak duration $\kappa \tau^+$, the higher is the threshold, the lower is the value of $n(\kappa \tau^+)$. Both these trends of variation result from the linear increase of the restoring drift  with the increasing height of the threshold $x_\text{th}$. Growing strength of the restoring drift due to the  active sensing of the flagellar length  causes stronger suppression of longer excursions beyond a given threshold and suppression of excursions beyond higher thresholds as well.

\subsection{Relevant lengthscales}
Now we look at the extreme excursions of the protrusion length about its mean  value $\ell_\text{ss}$ in the steady state. The quantities of immense interest are the extreme length the protrusion can grow or shorten to and the width of the range it scans within a finite duration of time \cite{guillet19,masoliver14,hartich19}. As shown in Fig.\ref{fig_len}(a), the maximum $\ell_\text{max}(t)=\ell_\text{ss}+x_\text{max}(t)$ and the minimum  $\ell_\text{min}(t)=\ell_\text{ss}-x_\text{min}(t)$  values that the flagellum with fluctuating length can attain and the range $\ell_\text{range}(t)$ which the tip can scan are time dependent random variables. {\it Chlamydomonas reinhardtii} assembles its flagellum at the beginning of G1 phase of the cell cylce and starts to disassemble at the end of this phase when the cell prepares itself for cell division. G1 phase lasts for 8-12 hours. By considering that initially the length of the flagellum $\ell_0=\ell_\text{ss}$ , in Fig.{\ref{fig_len}}(b) we have plotted the average maximum $\langle \ell_\text{max}(t) |0 \rangle-\ell_\text{ss}~(=\langle x_\text{max}(t) |0 \rangle )$  and minimum length $\langle \ell_\text{min}(t) |0 \rangle -\ell_\text{ss}~(=\langle x_\text{min}(t) |0 \rangle) $ to which the flagellum can grow or shorten to by fluctuations by using the expressions (2.1) of Table.\ref{OU_Table} (see the derivation in appendix D) during its lifetime. The range $\langle \ell_\text{range}(t) |0 \rangle-\ell_\text{ss}~(=\langle x_\text{range}(t) |0 \rangle )$ which the flagellar tip could scan during its lifetime of 8 hours for the flagellum of a wildtype cell is estimated by using the expression (2.2) (see the derivation in appendix D) and plotted as a function of time in Fig.\ref{fig_len}(c). The magnitude of the three quantities plotted in Fig.\ref{fig_len}(b-c)  increase monotonically with time. This observation is in accordance with the fundamental principle of extreme value statistics which states that  ``{\it longer the wait, bigger the fish you catch}''. Another interesting thing to note is that all these three quantities  characterizing the extremal excursions are proportional to $\sqrt{Dt}$ (see expressions (2.1) and (2.2) in Table.\ref{OU_Table}) just like the root mean square displacement in pure diffusion but none of these depend on $\kappa$ at all. Since it is solely because of diffusion that the particle tends to move away from the mean position and explore extremes, all these three time-dependent quantities vary with time as $\sqrt{Dt}$, the hall mark of diffusion. The expression (2.1) and (2.2) of Table.\ref{OU_Table} and  equation (\ref{def_kappa_D}) and (\ref{eq-ABCflag}) together also indicate that these lengthscales associated with extreme excursion are proportional to $\sqrt{\Gamma_r}$. By varying the amount of depolymerase which control $\Gamma_r$ at the flagellar tip, the following scaling relations 
\begin{eqnarray}
\langle \ell_{i}(t)|\ell_\text{ss},0 \rangle \propto (B t)^{1/2}=(1-\rho)\sqrt{ \Gamma_r t} \nonumber\\~~~~(i=\text{max,min,range})
\label{eq-ext-quant}
\end{eqnarray}
can be verified.

\begin{figure*}
\includegraphics[width=0.85\textwidth]{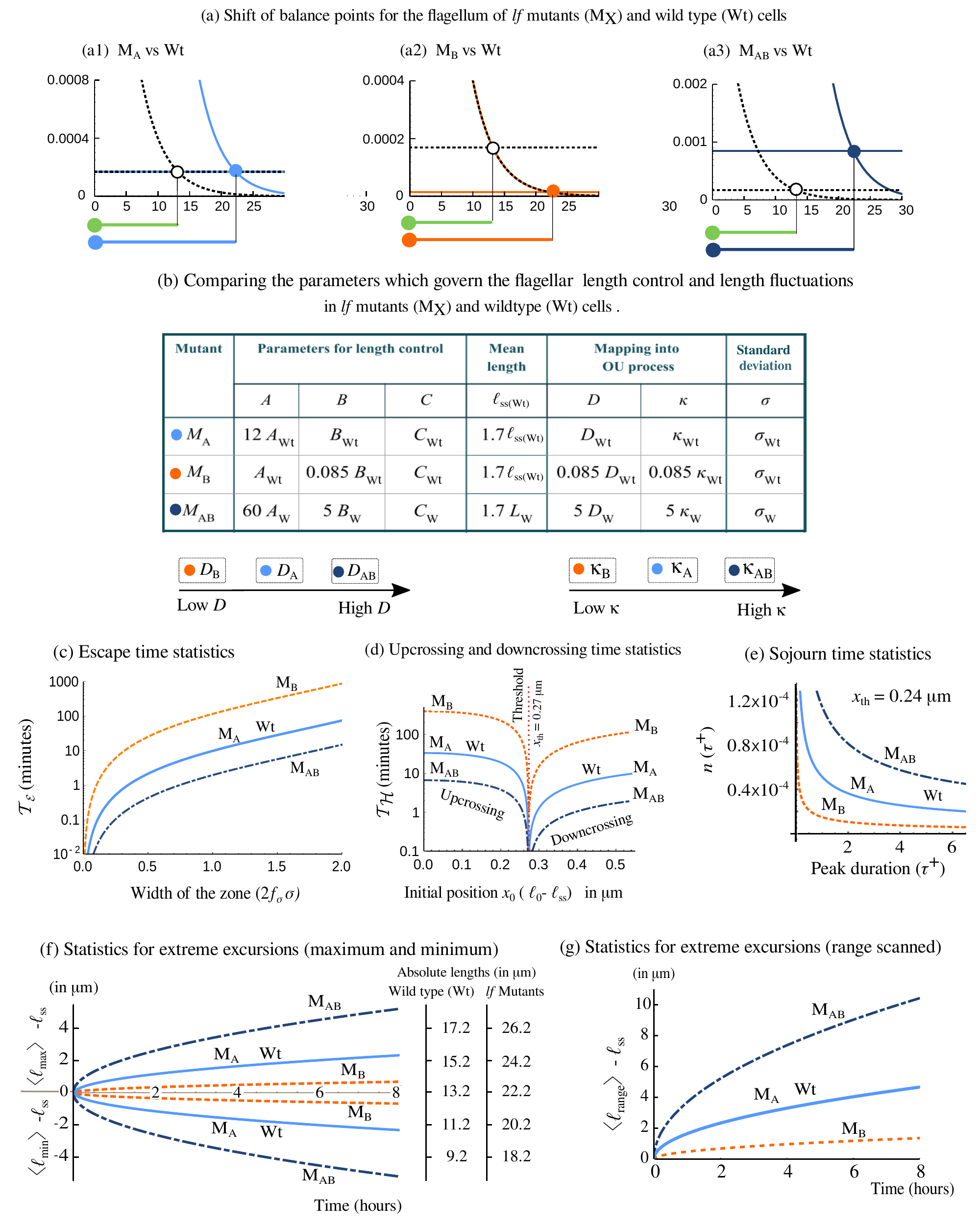}
\caption{Level crossing statistics for mutant cells with longer flagellum: (a) Three long mutants M$_\mathbb{X}$ are obtained by modifying either one or two of the three parameters $A_\text{Wt}$, $B_\text{Wt}$ and $C_\text{Wt}$ which govern the mean length $ \ell_\text{ss(Wt)}$ of the flagellum of the wild type cell in the steady state. (a1-a3) The assembly and the disassembly rates are plotted as a function of flagellar length and the location of the balance point for the wild type cell (dotted lines) and that for the $lf$ mutants (solid lines) are indicated in each plot. (b) In the table, the values of $A$, $B$ and $C$ are noted for each mutants and the corresponding values of $\kappa$, $D$ and $\sigma$ are noted. The magnitude of $\kappa$ and $D$ associated with the mutants and the wild type cell are arranged in increasing order.  (c-f) Statistics of level crossing quantities. (c) Mean escape time as a function of the width of the symmetric zone. (d) Mean upcrossing and downcrossing time as a function of initial length for hitting a given threshold. (e) Number density of peaks beyond a given threshold as a function of peak time or sojourn time. (f) Mean maximum and minimum length attained and (g) the average  range scanned  by the flagellum of different mutants as a function of time.}
\label{fig_mutants}
\end{figure*}
\subsection{Length fluctuations in mutants}
As indicated by eq.(\ref{lss}), a flagellum of longer or shorter mean length in the steady state can result from the alteration of the numerical value of one or more of the  intraflagellar parameters, like number density of IFT trains ($\rho$) or their flux ($J$) or velocity ($v$), association ($\Omega_e$) and dissociation ($\Gamma_r$) kinetics of monomers at the tip or the pool size ($n_\text{ss}$ or $n_\text{max}$), that might be caused by a mutation. Changes in these parameters change at least one of the three parameters $A,B$ or $C$ (see eq.(\ref{eq-ABCflag}) and eq.(\ref{lss})). With three {\it in silico} mutants M$_\mathbb{X}$, where $\mathbb{X}$ denotes the set of parameters which have been altered to produce a mutant, we demonstrate that even if the steady state mean length $\ell_\text{ss}$ of these $lf$ mutants are same, the level-crossing statistics of length fluctuations differ  quantitatively. In mutants M$_\text{AB}$ the values of the parameters $A$ and $B$ have been altered by multiplying $A_\text{Wt}$ and $B_\text{Wt}$ with different factors whereas in  mutants M$_\text{A}$ and M$_\text{B}$, parameters A and B are altered respectively. The remaining parameters in these three mutants M$_\text{A}$, M$_\text{B}$ and M$_\text{AB}$ are same as that in the wild type cell. For the values of $A_\text{Wt}$, $B_\text{Wt}$ and $C_\text{Wt}$ see eq.(\ref{eq-wt}).

We have designed these mutants  by altering the parameters  in such a way that the steady state length  $ \ell_{\text{ss}{(\mathbb{X})}}$ of all these mutants are same i.e, $ \ell_{\text{ss}{(\mathbb{X})}}$ is 1.7 times longer than the steady state length  $ \ell_{\text{ss}{\text{(Wt)}}}$ of the wild type flagellum ($ \ell_{\text{ss}{(\mathbb{X})}}=1.7~ \ell_{\text{ss}{(\text{Wt})}}$). However, we have the same $D/\kappa=2\sigma$ for them. In Fig.\ref{fig_mutants}(a), we have plotted the length-dependent assembly and disassembly rates as functions of the flagellar length for the wild type (Wt) cell as well as for these mutant cells. Note the shift in the location of the balance points of the mutants from that of the wild type cell. In the table shown in Fig.\ref{fig_mutants}(b), we list the values of the parameters $A,B$ and $C$ used for the plots displayed in Fig.\ref{fig_mutants}(a) for the three mutants. Instead of presenting the absolute values, we have listed the values of these parameters for the mutants as multiples (of fractions) of the corresponding parameters $A_\text{Wt}$, $B_\text{Wt}$ and $C_\text{Wt}$ for the wild type cell.  

We compare the statistics of level crossing quantities for these mutants these three mutants. The curves in Fig.\ref{fig_mutants}(c) indicate that the flagellum of fluctuating length of the mutant M$_\text{AB}$ takes the least time to escape a symmetric zone of a given width whereas the one belonging to the mutant M$_\text{AB}$ takes the maximum time for escape. The hitting time for hitting a given threshold either by upcrossing it (upcrossing time) or downcrossing it (downcrossing time) is plotted as a function of initial length in Fig.\ref{fig_mutants}(d). Again both these mean hitting times is minimum for the flagellum of mutant  M$_\text{AB}$ and maximum for the one belonging to the mutant M$_\text{B}$. The number density of peaks of a certain duration is comparatively higher for the mutant M$_\text{AB}$ whereas it is least for mutant M$_\text{B}$ as indicated by the curves in Fig.\ref{fig_mutants}(e). The average extreme lengths (Fig.\ref{fig_mutants}(f)) and the range scanned (Fig.\ref{fig_mutants}(g)) by the flagellum in a given duration of time is the highest for the mutant M$_\text{AB}$ and smallest for mutant M$_\text{B}$.

Even if the ratio $D/\kappa$ is same for all these three mutants used for the graphical plots, the  diffusion constants satisfy the inequalities  $D_\text{B}<D_\text{A}<D_\text{AB}$. This indicates the trends observed in the case of level crossing quantities for the flagellum of the mutants in Fig.\ref{fig_mutants}(c-g). For the mutant M$_\text{AB}$ with highest value of $D$, it is much easier for escaping a given zone or hitting a given threshold (Fig.\ref{fig_mutants}(c-d)). Hence, the mean escape time and upcrossing times is minimum. The drift also satisfy the inequalities  $\kappa_\text{B}<\kappa_\text{A}<\kappa_\text{AB}$. For the flagellum of the downcrossing time is also minimum because strongest $\kappa$ ensures quicker restoration towards the mean value (Fig.\ref{fig_mutants}(d)). Stronger $D$ for the flagellum of the mutant M$_\text{AB}$ is also responsible for highest number of peaks beyond a given threshold (Fig.\ref{fig_mutants}(e)) and also for the highest average maximum and minimum lengths and range scanned (Fig.\ref{fig_mutants}(f-g)). Higher value of $D$ makes the excursion easier beyond a given threshold and also drive the excursions towards the extremes more frequently. 

\section{Discussion: summary and conclusion}
{
In this section we highlight what has been achieved in this paper. More specifically, we answer the four questions (a)-(d) posed in the introduction (section I) of this paper; answers to the questions (b)-(d) are also new predictions that can, at least in principle, be tested experimentally. We also discuss the relevance of our work in prescribing stringent tests for the validity of competing theoretical models of flagellar length control. Finally, we explain how the generic theoretical framework developed here for analyzing level-crossing statistics of dynamic protrusions can also be adopted for similar analysis length fluctuations in other long cell appendages and filaments.

\subsection{Flagellar length fluctuations as OU process}
We begin by answering the question (a), namely the type of stochastic process that best describes the flagellar length fluctuations. 
In a recent paper, Bauer et al. \cite{bauer20} have reported experimental results on length fluctuations of individual flagellum, both for wild type cells and mutants bearing long flagellum. In both the cases they observed that the flagellar tip performs a confined random walk about its mean position in the steady state and  proposed that the stochastic kinetics of the flagellar tip is analogous to that of a Brownian particle attached to a spring. Hence, flagellar length fluctuations can be mappped onto {\it Ornstein-Uhlenbeck} (OU) process of the flagellar tip. 

Independently, in this paper, we have arrived at the same conclusion by analysing the length fluctuations of our model flagellum which grows and shortens by adding and removing precursors at its distal end. 
Our derivation, that starts from the master equations for the kinetics of the model, explicitly
shows how the two important parameters for OU process, namely $D$ and $\kappa$, depend on various parameters associated with intraflagellar processes, like the traffic properties of IFT trains, kinetics of underlying microtubule doublets and strength of the pool, which collectively control flagellar assembly. Since level-crossing statistics for general form of OU process is known, the trick of mapping of our problem for a temporally fluctuating flagellar length onto the OU process leads to enormous simplification of deriving the desired results. 

Our derivation also demonstrates that, in the steady state, in all models based on the {\it balance-point scenario} the kinetics of the flagellar tip gets mapped onto the OU process. However, since the  functional forms of $r^+(\ell)$ and $r^-(\ell)$ differ from one model to another, the functional dependence of $D$ and $\kappa$ on intraflagellar parameters will also be different from those in eq.(\ref{def_kappa_D}). Nevertheless, for all the models based on the balance-point scenario, the known results for OU process can be exploited for the derivation of the analytical expressions for the corresponding level-crossing statistics.

\subsection{Predictions on level-crossing statistics}

In this subsection we summarize the answers to the questions (b)-(d), which should also be treated as new theoretical predictions that can, in principle, be tested experimentally. By mapping flagellar length fluctuations onto the OU process, we then utilized the known results of OU process for obtaining the level crossing statistics for flagellar length fluctuations. 

First, we mention consistency of one of the results of our study with an experimentally observed fact. Altering the intraflagellar parameters we generated  a few model flagella of in-silico mutants with distinct features. Our study reveals how alteration in the intraflagellar parameters  $A,$ $B,$ and $C$ leads to change in  $\kappa$ and $D$ which results in quantitative differences between the fluctuation statistics of the wildtype and the mutants as well as those of the different mutants. Our analysis corroborates the conclusions drawn by Bauer et al. \cite{bauer20} that significant quantitative differences in the fluctuation statistics is possible even if the mean lengths in the steady state are comparable. 

Next we highlight the testable major new predictions on the level crossing statistics: (i) Mean escape time $\mathcal{T}_\mathcal{E}$ goes as $x^2_\text{th}$ for $x_\text{th}<\sigma$ (eq.(\ref{eq-OnlyDdep1})) and as $x^4_\text{th}$ for $x_\text{th}>\sigma$ (eq.(\ref{eq-OnlyDdep2})). Besides, the nonlinear dependence on various measurable quantities like speed of IFT train $v$ or effective disassembly rate $B$ can also be measured using wild type cells and mutants (eq.(\ref{eq-OnlyDdep1}-\ref{eq-OnlyDdep2})). (ii)  Trends like  the asymmetries in the upcrossing and downcrossing times for hitting a threshold and (iii) sharp fall of number of peaks with increasing peak duration beyond a given threshold  are straightforward to verify. (iv) Square root dependence of the quantities of extreme excursions  on time (eq.(\ref{eq-OnlyDdep1})) are easy to check by analysing the trajectory of flageller with time of wild type flagellum and the square root dependence on the effective disassembly rate (eq.\ref{eq-OnlyDdep2}) can be verified with different mutants with different effective rates for disassembly (the amount of depolymerase at the tip are a measure for disassembly rate). 

}

\subsection{Length fluctuations as tools for testing different length control models}
As stated earlier, several different models of flagellar length control have been reported in the literature \cite{ludington15}. Only a few of these considered some aspects of the fluctuations and noise. For example, a tip to base gradient of RanGTP could control the injection of IFT trains which, in turn, leads to the falling rate of elongation with the increasing flagellar length. Based on this, Bressloff and Karamchend proposed a stochastic model for flagellar length control \cite{bressloff18}. The key feature of their model is that it accounts for the qualitative features of the time series data of IFT injections reported in ref.\cite{ludington13}. 

Alternatively, the flux of IFT trains can also be regulated as a function of flagellar length, for example, by the ciliary current \cite{besschetnova10}, a gradient of  kinesin motors diffusing  from the flagellar tip to the base \cite{hendel18} or due to the  limited availability of IFT trains \cite{marshall01}. In contrast to the above mentioned scenarios, our model is based on the idea of differential loading of flagellar  precursor proteins on the IFT trains \cite{patra20,wren13}. In our model, it is the decreasing amount of cargo per IFT train which causes  the  effective elongation rate of the flagellum  to fall with the increasing flagellar length and the cell senses the growing length based on a time-of-flight mechanism \cite{patra20}. 

The common theme in all the models, including ours, where flux of the IFT trains or the cargo carried by them is regulated is that the assembly rate falls with the growing flagellar length whereas the disassembly rate remains constant \cite{hendel18, bressloff18, patra20, marshall01}. But,  the rates of both the assembly and disassembly are assumed to be length-dependent in the models proposed by Fai et al  \textit{Chlamydomonas} \cite{fai19} and by  McInally et al. \cite{mcinally19}. 
Since all these competing models can account for the time-dependence of the mean flagellar length, additional tests are needed to rule out at least some of these models. 

Bauer et al \cite{bauer20} have indicated that in near future it may be possible to test the validity of the models of flagellar length control by comparing the nature of temporal fluctuations predicted by these models with the corresponding experimental data. Our  generalised framework developed here in terms of level-crossing statistics is likely to be useful in this endeavour. In this paper we have reported the level-crossing statistics in our model of flagellar length control. In principle, it should be possible to calculate the same quantities also for other competing models mentioned above.  Some of the models that successfully account for the time-dependence of average length of flagella may still have to be discarded if they fail the test of reproducing the experimentally measured data on level-crossing statistics when such data experimental become available.

\subsection{Relevance from a broader perspective}

A model flagellum can attain a stationary (i.e., time-independent) mean length, in principle, if (i) the assembly rate decreases with increasing length while the disassembly rates is independent of length, or (ii) the disassembly rate increases with increasing length while the assembly rate remains constant, or (iii) both these phenomena occur. These three special cases are referred as (i) assembly controlled, (ii) disassembly controlled and (iii) assembly-disassembly controlled respectively. In each of these cases, the cell can actively `sense' the present length of its flagellum and adjust the rates of assembly or disassembly, or both, resulting in length-dependent rates. In this paper we have considered an assembly-controlled model, a scenario that is one of the oldest and most popular. 
It should be straightforward to carry out similar study of level-crossing statistics also for the disassembly controlled and assembly-disassembly controlled models which differ from each other in terms of the expressions for 
$r^+(\ell)$ and $r^-(\ell)$. 

In the past, models have been developed for understanding the mechanisms of length control of long protrusions like (i) cilia and flagella \cite{ludington15,patra20,fai19}, (ii) stereocilia \cite{orly15,prost07}, (iii) microvilli \cite{gov06,prost07}, (iv) axon \cite{folz19} and cytoskeleton filaments  like (i) microtubule \cite{rank18,melbinger12,govindan08,kuan13,klein05,johann12} and (ii) actin \cite{mohapatra15}. Except for a few papers, including ref.\cite{patra20,bauer20,gov06}, all the works on length control of cell protrusions  and other linear subcellular structures, so far have investigated the time-dependence of the mean lengths. Inspired by the very recent developments in the studies of flagellar length fluctuations, similar studies of other protrusions may also begin.  

The spirit of our investigation of the fluctuations of cell protrusions has some overlap with that in ref.\cite{guillet19}. In the latter the authors have analyzed the extreme value statistics of molecular motors that  walk on filamentous tracks by consuming a chemical fuel.  In our work the tip of the cell protrusion performs a biased random walk while in ref.\cite{guillet19} the movement of the motor is modelled as a biased random walk.  But, there is a fundamental difference between our model and that in ref.\cite{guillet19}. In the latter, the bias results in an average drifting of the molecular motor away from its initial position. In contrast, in our model, the bias causes an  effective drifting of the random walker towards the origin which corresponds to a tendency of restoration of the tip of the protrusion to its mean position.

{\bf Acknowledgement :} SP acknowledges support from IIT Kanpur through a teaching assistantship.  D.C. acknowledges support from SERB (India) through a J.C. Bose National Fellowship. 

\appendix

\section{Mapping onto Ornstein-Uhlenbeck process}
\label{map_to_ou}

Utilizing the well known correspondence between Fokker-Planck and Langevin equations, we obtain the Langevin equation \cite{kampen10,gardiner09} 
\begin{equation}
dy=R_{-}(y)dt+\frac{1}{\sqrt{N}} \sqrt{R_{+}(y)}dW(t) 
\label{eq-Ylangevin-gp}
\end{equation}
for the Fokker-Planck equation (\ref{fpe-eq-gp}) that describes the stochastic evolution of the position of the tip, where $W(t)$ is the Gaussian white noise (a Wiener process)  with $dW(t)$ distributed according to a Gaussian process with mean and covariance given by
\begin{eqnarray}
\langle dW(t) \rangle &=& 0 ~~\text{(mean)} \nonumber\\  \langle dW(t) dW(s) \rangle &=& \delta(t-s)~dt~ ds ~~\text{(covariance)} ~.
\label{eq-GWN}
\end{eqnarray}
The Langevin equation (\ref{eq-Ylangevin-gp}) describing the stochastic evolution of the protrusion length involves Gaussian like fluctuations of order $1/\sqrt{N}$ about the deterministic trajectory. As we are interested in studying the properties of steady state length fluctuations, we make a change of variable from $y$ to $x$  by defining
\begin{eqnarray}
y-y_\text{ss}=\frac{x}{\sqrt{N}} \Rightarrow y=y_\text{ss}+\frac{x}{\sqrt{N}} 
\end{eqnarray}
Substituting this in equation (\ref{eq-Ylangevin-gp}) we get
\begin{widetext}
\begin{eqnarray}
 d(y_\text{ss}+\frac{x}{\sqrt{N}}) & = & R_- (y_\text{ss}+\frac{x}{\sqrt{N}})dt+\frac{1}{\sqrt{N}}\sqrt{R_+(y_\text{ss}+\frac{x}{\sqrt{N}})}dW(t) \nonumber\\
 \Rightarrow \frac{dx}{\sqrt{N}} & = & \bigg{(} R_- (y_\text{ss})+\frac{x}{\sqrt{N}}R_+' (y_\text{ss})  \bigg{)} dt+\frac{1}{\sqrt{N}}\sqrt{ \bigg{(} R_+ (y_\text{ss})+\frac{x}{\sqrt{N}}R_+' (y_\text{ss}) \bigg{)} }dW(t)
\end{eqnarray} 
\end{widetext}
where prime in $R_{\pm}'$ indicates derivative of $R_{\pm}(y)$ with respect to $y$. 
Using the solution of the rate equation (\ref{lss}) in steady state, we get $R_-(y_\text{ss})=0$. Neglecting the term $R_+'(y_\text{ss})$ inside the square root we get
\begin{eqnarray}
\Rightarrow \frac{dx}{\sqrt{N}} &=& \frac{x}{\sqrt{N}}R_+' (y_\text{ss}) dt+\frac{1}{\sqrt{N}}\sqrt{R_+ (y_\text{ss})}dW(t) \nonumber \\
 \Rightarrow  dx &=& R_+' (y_\text{ss}) x~ dt+\sqrt{R_+ (y_\text{ss})}dW(t)
\end{eqnarray}
which is the transformed stochastic differential equation
\begin{eqnarray}
dx={\cal R}_{-}(x)~dt+\sqrt{{\cal R}_{+}(x)}~dW(t)~. 
\label{transformed_SDE-gp}
\end{eqnarray}
The corresponding Fokker-Planck equation is 
\begin{eqnarray}
\frac{\partial p(x,t)}{\partial t}=\frac{\partial }{\partial x} \bigg{[} \kappa  x \ p(x,t)\bigg{]}+\frac{D}{2}\frac{\partial^2 p(x,t)}{\partial x^2}
\end{eqnarray}
where 
\begin{equation}
\kappa= - R'_{-}(y_\text{ss}) ~ \&  ~ D=R_{+}(y_\text{ss})~.
\label{def_kappa_D}
\end{equation}
\section{Statistics of passage times}
 
As is well known, calculation of mean first passage times are normally more convenient if one uses backward Fokker-Planck equation, rather than the forward Fokker-Planck equation given by equation (\ref{FPE-OU-gp}) \cite{gillespie13}. For the generic model under our consideration, the backward Fokker-Planck equation is given by 
{\small
\begin{eqnarray}
-\frac{\partial p(x,t|x_0,t_0) }{\partial t_0}=-\kappa  x_0 \frac{\partial p(x,t|x_0,t_0)}{\partial x_0} +\frac{D}{2}\frac{\partial^2 p(x,t|x_0,t_0)}{\partial x_0^2} \nonumber\\
\label{BFPE-def}
\end{eqnarray} }
As both the drift $\kappa x_0$ and the diffusion $D$ are not dependent on time $t$ and $t_0$ explicitly, $p(x,t|x_0,t_0)$ depends on time only through the difference $t-t_0$. Therefore, the evolving protrusion length is considered to be temporally homogenous system  for which the backward Fokker-Planck equation is given by
{\small
\begin{eqnarray}
\frac{\partial p(x,t|x_0,t_0) }{\partial t}=-\kappa  x_0 \frac{\partial p(x,t|x_0,t_0)}{\partial x_0} +\frac{D}{2}\frac{\partial^2 p(x,t|x_0,t_0)}{\partial x_0^2} 
\label{bfpe}
\end{eqnarray}    
}
as by chain rule $\partial_{t_0}=-\partial_{t}$.

The probability that the particle located at $x_0$ at time $t_0$ escapes the safe zone, bounded by the upper ($x_U$) and the lower ($x_L$) thresholds, for the first time at time $t$ is given by
\begin{eqnarray}
\mathcal{E}(t;x_U,x_L|\underline{x_0,t_0})=1-\underbrace{\int_{x_U}^{x_L} p(x,t|x_0,t_0)dx}_{\substack{\text{probability that the particle}\\ \text{is lying between $x_U$ and $x_L$}}}\nonumber\\
\label{esc-prob}
\end{eqnarray}

The probability of hitting a threshold $x_{th}$ is closely related to the escape probability defined in equation (\ref{esc-prob}). Let $\mathcal{H}(t;x_{th}|\underline{x_0,t_0})$ denote the probability of hitting the threshold $x_{th}$ at time $t$, given that the particle was at $x_0$ at time $t_0$. In case $x_{th}>x_0$ , the hitting (or upcrossing) probability is given by
\begin{eqnarray}
\mathcal{H}_{U}(t;x_{th}|\underline{x_0,t_0})=\mathcal{E}(t;x_U=x_{th},x_L=-\infty|\underline{x_0,t_0})
\label{upc-prob}
\end{eqnarray} 
which is like escaping a semi-infinite interval ($-\infty,x_{th}$).
On the other hand, if $x_{th}<x_{0}$,the hitting (or downcrossing) probability is given by
\begin{eqnarray}
\mathcal{H}_{D}(t;x_{th}|\underline{x_0,t_0})=\mathcal{E}(t;x_U=\infty,x_L=x_{th}|\underline{x_0,t_0})
\label{down-prob}
\end{eqnarray} 
which is like escaping a semi-infinite interval ($x_{th},\infty$).
In the following subsections, we will derive the expressions for the mean exit time from the region bounded by two thresholds and for the mean hitting time to a given threshold $x_{th}$.

\subsection{Escaping the  zone bounded by \\ an upper and a lower threshold}
\label{app-mfpt}
The mean escape time taken to escape the zone bounded by $x_U$ and $x_L$  respectively  is denoted by
\begin{equation}
\mathcal{T}_{\mathcal{E}}(x_U,x_L|\underline{x_0,t_0})=\int t~ p_{\mathcal{E}}(t,x_U,x_L|\underline{x_0,t_0})dt
\label{met-def}
\end{equation} 
where $p_{\mathcal{E}}(t,x_U,x_L|\underline{x_0,t_0})$ is the {\it pdf} corresponding to the escape probability $\mathcal{E}(t;x_U,x_L|\underline{x_0,t_0})$. 

The escape probability $\mathcal{E}(t;x_U,x_L|\underline{x_0,t_0})$ is introduced in equation (\ref{esc-prob}). For our convinience, we denote it simply by $\mathcal{E}(x_0,t)$ . Integrating the backward FPE (\ref{bfpe}) with respect to final position $x$ (as done while defining the escape probability in equation (\ref{bfpe})) it can be checked that the escape probability $\mathcal{E}(x_0,t)$ obeys the backward FPE i.e, 
{\small
\begin{eqnarray}
\frac{\partial  \mathcal{E}({x_0,t})}{\partial t}=-\kappa x_0 \frac{\partial [\mathcal{E}({x_0,t})}{\partial x_0}+\frac{D}{2}\frac{\partial^2 \mathcal{E}({x_0,t})}{\partial x_0^2}  
\label{esc-bfpe}
\end{eqnarray} }
subjected to the initial condition
\begin{equation}
\mathcal{E}({x_0,t_0})=0
\label{esc-ic}
\end{equation}
and the following boundary conditions
\begin{equation}
\mathcal{E}({x_U,t})=1 ~ \text{and} ~ \mathcal{E}({x_L,t})=1
\label{esc-bc}
\end{equation}
The initial condition (equation (\ref{esc-ic})) indicates that initially $x_0$ lies between $x_U$ and $x_L$ and the boundary condition (equation (\ref{esc-bc})) indicates whenever the length hits either of the thresholds, it successfully exits the zone. 

Using the Laplace transform 
\begin{equation}
\tilde{\mathcal{E}}(x_0,s)=\int_0^\infty e^{-st} \mathcal{E}({x_0,t})dt
\end{equation}
the backward FPE (equation (\ref{esc-bfpe})) gets converted to an ordinary differential equation 
\begin{eqnarray}
-\kappa x_0 \frac{d \mathcal{ \tilde{\mathcal{E}}}({x_0,s})}{dx_0}+\frac{D}{2}\frac{d^2 \tilde{\mathcal{E}}({x_0,s})}{dx_0^2}  =s \tilde{\mathcal{E}}({x_0,s})-1 
\label{esc-ode}
\end{eqnarray} 
with boundary conditions 
\begin{equation}
\mathcal{ \tilde{\mathcal{E}}}({x_L,s})=\mathcal{ \tilde{\mathcal{E}}}({x_U,s})=1/s
\label{esc-ode-bc}
\end{equation}
Now onwards, for our convinience, we will denote the corresponding {\it pdf} $p_{\mathcal{E}}(t,x_U,x_L|\underline{x_0,t_0})$  by a simpler notation $p_{\mathcal{E}}(x_0,t)$. The moments of the exit time are given by 
\begin{equation}
\mathcal{T}^n_{\mathcal{E}}(x_U,x_L|\underline{x_0,t_0})=\int_0^\infty t^n p_{\mathcal{E}}(x_0,t)dt
\end{equation}
where the zeroth moment corresponding to $n=0$ is $\mathcal{T}^0_{\mathcal{E}}=1$ and the first moment corresponding to $n=1$ is the mean exit time $\mathcal{T}_{\mathcal{E}}(x_U,x_L|\underline{x_0,t_0})$. Let us denote the moments of exit time $\mathcal{T}^n_{\mathcal{E}}(x_U,x_L|\underline{x_0,t_0})$ simply by $\mathcal{T}^n_{\mathcal{E}}(x_0)$. If $\tilde{p}_{\mathcal{E}}(x_0,s)$  is the Laplace transform of the {\it pdf} ${p}_{\mathcal{E}}(x,t)$, the moments of escape time are given by
\begin{equation}
\mathcal{T}^n_{\mathcal{E}}(x_0)=(-1)^n \frac{\partial^n }{\partial s^n}\tilde{p}_{\mathcal{E}}(x_0,s) \bigg{|}_{s=0}
\end{equation}
Therefore, $\tilde{p}_{\mathcal{E}}(x_0,s)$ can be expanded and its series form is given by
the Laplace transform $\tilde{p}_{\mathcal{E}}(x_0,s)$ can be written as a power series of the Laplace variable $s$ 
\begin{equation}
\tilde{p}_{\mathcal{E}}(x_0,s)= \sum_{n=0}^{\infty} \frac{(-1)^n}{n!} \mathcal{T}_{\mathcal{E}}^n{(x_0)s^n}
\label{p-esc-series}
\end{equation}
provided all the moments of the escape time exists.

$p_{\mathcal{E}}(x_0,t)$ is obtained from the probability $\mathcal{E}(x_0,t)$ by taking  derivate with respect to $t$
\begin{equation}
p_{\mathcal{E}}(x_0,t)=\frac{\partial \mathcal{E}(x_0,t)}{\partial t}~.
\label{pdf-prob}
\end{equation}
Taking into account the fact that $\mathcal{E}(x_0,0)=0$ which indicates that initially it is impossible to escape the safe zone, it can be shown that the relation between the Laplace transforms $\tilde{\mathcal{E}}(x_0,s)$ and $\tilde{p}_{\mathcal{E}}(x_0,s)$ is
\begin{equation}
\tilde{p}_{\mathcal{E}}(x_0,s)=s~\tilde{\mathcal{E}}(x_0,s)
\label{pdf-prob-lt}
\end{equation}
which is obtained by taking the Laplace transforms of the terms on both side of the equation (\ref{pdf-prob}). From equation (\ref{p-esc-series}) and(\ref{pdf-prob-lt}), we can see that the 
\begin{eqnarray}
\tilde{\mathcal{E}}(x_0,s)=\frac{1}{s} \sum_{n=0}^{\infty} \frac{(-1)^n}{n!} \mathcal{T}_{\mathcal{E}}^n{(x_0)s^{n}} \nonumber\\
=\frac{1}{s} \bigg{(} 1-s \mathcal{T}_{\mathcal{E}}{(x_0)} +\sum_{n=2}^{\infty} \frac{(-1)^n}{n!} \mathcal{T}_{\mathcal{E}}^n{(x_0)s^{n}}\bigg{)}
\label{esc-lt-series}
\end{eqnarray} 
On rearranging the above equation
\begin{eqnarray}
\frac{1}{s}-\tilde{\mathcal{E}}(x_0,s)=\mathcal{T}_{\mathcal{E}}{(x_0)} - \sum_{n=2}^{\infty} \frac{(-1)^n}{n!} \mathcal{T}_{\mathcal{E}}^n{(x_0)s^{n-1}} 
\end{eqnarray} 
and taking the limit $s \to 0$ we get
\begin{equation}
\lim_{s \to 0} \bigg{[}\frac{1}{s}-\tilde{\mathcal{E}}(x_0,s)\bigg{]}=\mathcal{T}_{\mathcal{E}}{(x_0)} ~ .
\label{limit-esc-met}
\end{equation}
Plugging this relation mentioned in equation (\ref{limit-esc-met}) into the backward Fokker-Planck equation (\ref{esc-bfpe}), we get 
\begin{widetext}
\begin{eqnarray}
\lim_{s \to 0} \bigg{\{}-\kappa x_0 \frac{d }{dx_0} \bigg{[} \frac{1}{s} - \mathcal{T}_\mathcal{E} (x_0)  \bigg{]}+\frac{D}{2}\frac{d^2 }{dx_0^2} \bigg{[} \frac{1}{s} - \mathcal{T}_\mathcal{E} (x_0)  \bigg{]}        \bigg{\}}=\lim_{s \to 0} \bigg{\{} s \bigg{[} \frac{1}{s} - \mathcal{T}_\mathcal{E} (x_0)  \bigg{]}-1 \bigg{\}}
\end{eqnarray}
\end{widetext}
which simplifies to 
\begin{equation}
-\kappa x_0 \frac{d  \mathcal{T}_{\mathcal{E}}(x_0) }{dx_0}+\frac{D}{2}\frac{d^2 \mathcal{T}_{\mathcal{E}}(x_0)}{dx_0^2}=-1 ~ .
\label{mean-esc-ode}
\end{equation}
On rearranging the corresponding boundary condition (\ref{esc-ode-bc}) and taking limits
\begin{equation}
\lim_{s \to 0} \bigg{[}\frac{1}{s}-\tilde{\mathcal{E}}(x_U) \bigg{]}=
\lim_{s \to 0} \bigg{[} \frac{1}{s}-\tilde{\mathcal{E}}(x_L) \bigg{]}=0
\end{equation}
we get the boundary condition 
\begin{equation}
\mathcal{T}_{\mathcal{E}}(x_U)=\mathcal{T}_{\mathcal{E}}(x_L)=0.
\end{equation}
for the ordinary differential equation (\ref{mean-esc-ode}). On solving it, we get the expression for the mean escape time $\mathcal{T}_{\mathcal{E}}(x_U,x_L|\underline{x_0,t_0})$ which reads as
\begin{widetext}
{\small
\begin{eqnarray}
\mathcal{T}_{\mathcal{E}}(x_U,x_L|\underline{x_0,t_0})=
{\bigg{[}D \left(\text{erfi}\left(\frac{\sqrt{\kappa } x_L}{\sqrt{D}}\right)-\text{erfi}\left(\frac{\sqrt{\kappa } x_U}{\sqrt{D}}\right)\right)\bigg{]}}^{-1} \times
 \bigg{[} x_0^2 \left(\text{erfi}\left(\frac{\sqrt{\kappa } x_U}{\sqrt{D}}\right)-\text{erfi}\left(\frac{\sqrt{\kappa } x_L}{\sqrt{D}}\right)\right) \, _2F_2\left(1,1;\frac{3}{2},2;\frac{\kappa  x_0^2}{D}\right)\nonumber\\+x_L^2 \left(\text{erfi}\left(\frac{\sqrt{\kappa } x_0}{\sqrt{D}}\right)-\text{erfi}\left(\frac{\sqrt{\kappa } x_U}{\sqrt{D}}\right)\right) \, _2F_2\left(1,1;\frac{3}{2},2;\frac{\kappa  x_L^2}{D}\right)+x_U^2 \left(\text{erfi}\left(\frac{\sqrt{\kappa } x_L}{\sqrt{D}}\right)-\text{erfi}\left(\frac{\sqrt{\kappa } x_0}{\sqrt{D}}\right)\right) \, _2F_2\left(1,1;\frac{3}{2},2;\frac{\kappa  x_U^2}{D}\right)\bigg{]} \nonumber\\
\label{met-ode-sol-app}
\end{eqnarray}
}
\end{widetext}
 where the function $\text{erfi}(z)$  is the imaginary error function \cite{abramowitz72} whose series about $z=0$ is given by 
\begin{equation}
\text{erfi}(z) = \frac{1}{\sqrt{\pi}} \left( 2z+\frac{2}{3}z^3+\frac{1}{5}z^5+\frac{1}{21}z^7+... \right)
\label{erfi-exp}
\end{equation}
and $_2F_2\left(a_1,a_2;b_1,b_2;z \right)$ is the generalized hypergeometric function \cite{abramowitz72} whose series about $z=0$ is given by
\begin{equation}
_2F_2\left(a_1,a_2;b_1,b_2;z \right)= \sum_{j=0}^{\infty} \frac{(a_1)_j (a_2)_j}{(b_1)_j (b_2)_j} \frac{z^j}{j !}
\label{gen-hypg-exp}
\end{equation} 

\subsection{Hitting the thresholds}
\label{app-mht}
If there is a single threshold ($x_{th}$) of interest, the appropriate quantity is the mean  hitting time $\mathcal{T}(x_{th}|\underline{x_0,t_0})$ which is the mean time taken to hit the threshold  $x_{th}$ for the first time if initially its length is $x_0$. The mean hitting time is given by
\begin{equation}
\mathcal{T}_{\mathcal{H}}(x_{th}|\underline{x_0,t_0})=\int_0^\infty t ~ p_\mathcal{H}(t;x_{th}|\underline{x_0,t_0})dt
\label{eq-MeanExit10}
\end{equation}
where $p_\mathcal{H}(t;x_{th}|\underline{x_0,t_0})$ is the  {\it pdf} associated with the hitting probability 
$\mathcal{H}(t;x_{th}|\underline{x_0,t_0})$.

By integrating the backward FPE (\ref{bfpe}) with respect to the final position and using the defination of hitting probability  given in equation (\ref{upc-prob}) or (\ref{down-prob}), we can show  that $\mathcal{H}(t;x_{th}|\underline{x_0,t_0})$ (simply written as $\mathcal{H}(x_0,t)$) obeys the backward FPE: 
\begin{eqnarray}
\frac{\partial  \mathcal{H}(x_0,t)}{\partial t}=-\kappa x\frac{\partial \mathcal{H}(x_0,t)}{\partial x_0}+\frac{D}{2}\frac{\partial^2 \mathcal{H}(x_0,t)}{\partial x_0^2}
\label{hp-bfpe}
\end{eqnarray} 
with initial condition
\begin{equation}
\mathcal{H}_(x_0,0)=0 
\label{hp-ic}
\end{equation}
and boundary condition given by
\begin{equation}
\mathcal{H}(x_{th},t)=1 .
\label{hp-bc}
\end{equation}

We rescale $t$ by multiplying it with $\kappa$. So, replacing the $t$ with the rescaled time $t_{r}$ which is given by
\begin{equation}
t_r=\kappa t ~.
\end{equation} 
we transform the partial differential equation 
\begin{equation}
\frac{\partial  \mathcal{H}(x_0,t_r)}{\partial t^r}=- x\frac{\partial \mathcal{H}(x_0,t_r)}{\partial x_0}+\frac{D}{2 \kappa}\frac{\partial^2 \mathcal{H}(x_0,t_r)}{\partial x_0^2}
\label{hp-tr-bfpe}
\end{equation}
The laplace transform of the hitting probability is given by 
\begin{equation}
\tilde{\mathcal{H}}(x_0,s)=\int_0^{\infty} e^{-st_r} \mathcal{H}(x_0,t_r) dt_r
\end{equation}
and using this Laplace transform, we convert the partial differential equation (equation (\ref{hp-bfpe})) into an ordinary differential equation given by
\begin{eqnarray}
- x\frac{d \tilde{\mathcal{H}}(x_0,s)}{dx}+
\frac{D}{2\kappa}\frac{d^2 \tilde{\mathcal{H}}(x_0,s)}{dx^2}={s} \tilde{\mathcal{H}}(x_0,s)
\label{hp-ode}
\end{eqnarray}
which is subjected to the boundary condition
\begin{equation}
\tilde{\mathcal{H}}(x_{th},s)=\frac{1}{s} ~.
\label{hp-ode-bc}
\end{equation}
Using the change of variables
\begin{equation}
z=\frac{\kappa}{D}x_0^2
\end{equation}
one can check that equation (\ref{hp-ode}) can be recasted to the following second order ODE
\begin{equation}
z\frac{d^2 \tilde{\mathcal{H}}(z,s)}{dz^2}+\bigg{(} \frac{1}{2} -z \bigg{)} \frac{d \tilde{\mathcal{H}}(z,s)}{dz}-\frac{s}{2}
\tilde{\mathcal{H}}(z,s)=0
\end{equation}
whose general solution is a linear combination of Kummer functions. Hence the general solution of the equation (\ref{hp-ode}) is given by
\begin{eqnarray}
\tilde{\mathcal{H}}(x_0,s)&=& C_1 F(s/2,1/2, \kappa x_0^2/D )\nonumber\\
&+& C_2 U(s/2,1/2, \kappa x_0^2/D )
\end{eqnarray}
Now let us have a look at the asymptotic behaviour of the Kummer functions. In the limit $x_0 \to \pm \infty$

\begin{eqnarray}
\lim_{x_0 \to \pm \infty} F(s/2,1/2, \kappa x_0^2/D )  \nonumber\\=\frac{\Gamma 
(1/2)}{\Gamma(s/2)} x_0^{s-1} e^{{\kappa x_0^2/D}} [1+\mathcal{O}(1/x_0^2)] \to \infty
\end{eqnarray} 
so, it can serve to be the solution when $x_0$ is bounded i.e, $x_0^2<x_{th}^2$. On the other hand,

\begin{eqnarray}
\lim_{x_0 \to \pm \infty} U(s/2,1/2, \kappa x_0^2/D )=|x_0|^{-s}[1+\mathcal{O}(1/x_0^2)] \to 0 \nonumber\\
\end{eqnarray} 
and this indicates that it can serve as the solution when $x_0$ remains unbounded, i.e, $x_0^2> x^2_{th}$.

$C_1$ and $C_2$ can be evaluated using the boundary condition  given by equation (\ref{hp-ode-bc}) and collectively, the unique solution of equation (\ref{hp-ode}) is given by
\begin{eqnarray}
\tilde{\mathcal{H}}(x_0,s)=  \begin{cases}       \frac{F(s/2,1/2,\kappa x_0^2/D )}{s F(s/2,1/2, \kappa x_{th}^2/D )}  \ \ \text{when $|x_0|<|x_{th}|$} \\ 
\\
\frac{U(s/2,1/2, \kappa x_0^2/D )}{s U(s/2,1/2, \kappa x_{th}^2/D )} \ \ \text{when $|x_0|>|x_{th}|$} 
\end{cases}
\nonumber\\
\label{hp_sol}
\end{eqnarray}

Using the arguments used to derive the relation stated in equation (\ref{limit-esc-met}) , we can arrive at the following  
\begin{equation}
\mathcal{T}^r_{\mathcal{H}}(x_{th}|\underline{x_0,t_0})=\lim_{s \to 0} \bigg{[} \frac{1}{s} - \tilde{\mathcal{H}}(x_0) \bigg{]}
\label{mean_tl}
\end{equation}
where $\mathcal{T}^r_{\mathcal{H}}(x_{th}|\underline{x_0,t_0})$ ( simply denoted by $\mathcal{T}^r_{\mathcal{H}}(x_0)$) is the rescaled mean hitting time.

For $x_0<x_{th}$, the rescaled mean hitting (upcrossing) time $\mathcal{T}^r_{\mathcal{HU}}(x_0)$  for hitting the threshold is
\begin{eqnarray}
\mathcal{T}^r_{\mathcal{H}}(x_0)=\mathcal{T}^r_{\mathcal{HU}}(x_0)= \lim_{s \to 0} \bigg{[}\frac{1}{s}- \frac{ F(s/2,1/2,\kappa x_0^2/D) }{s F(s/2,1/2,\kappa x_{th}^2/D)} \bigg{]}. \nonumber\\
\end{eqnarray}

Plugging the following relation in the above equation
\begin{eqnarray}
 F(s/2,1/2,\kappa x^2/D)\nonumber\\= 1+{s} \int_0^{\kappa x^2/D}F(1,3/2,z)dz + \mathcal{O}[s^2]
\end{eqnarray} 
we get 
\begin{widetext}
\begin{eqnarray}
\mathcal{T}_{\mathcal{HU}}(x_{th}|\underline{x_0,t_0})=\mathcal{T}_{\mathcal{HU}}(x_0)=\frac{1}{\kappa} \mathcal{T}^r_{\mathcal{HU}}(x_0)= \frac{1}{\kappa} \int_{\kappa x_0^2 /D}^{\kappa x_{th}^2 /D} F(1,3/2,z)dz=\frac{\sqrt{\pi}}{\kappa} \int_{\sqrt{{\kappa}} |x_0|/ \sqrt{D} }^{\sqrt{{\kappa}} |x_{th}|/ \sqrt{D}} e^{y^2} \text{erf} (y) dy \nonumber\\
=\frac{1}{D} \bigg{[}{  x_{\text{th}}^2 \, _2F_2\left(1,1;\frac{3}{2},2;\frac{\kappa  x_{\text{th}}^2}{D}\right)- x_0^2 \, _2F_2\left(1,1;\frac{3}{2},2;\frac{\kappa  x_0^2}{{D}}\right)}\bigg{]} 
\end{eqnarray}

For $x>x_{th}$, the rescaled mean hitting (downcrossing) $\mathcal{T}^r_{\mathcal{HD}}(x_0)$ time for hitting the threshold is
\begin{eqnarray}
\mathcal{T}^r_{\mathcal{H}}(x_0)=\mathcal{T}^r_{\mathcal{HD}}(x_0)= \lim_{s \to 0} \bigg{[}\frac{1}{s}- \frac{ U(s/2,1/2,\kappa x_0^2/D) }{s U(s/2,1/2,\kappa x_{th}^2/D)} \bigg{]}.\nonumber\\
\end{eqnarray}

Plugging the following relation in the above equation
\begin{small}
\begin{eqnarray}
U(s/2,1/2, \kappa  x^2/D)= 1-\frac{s}{2} \bigg{[} \psi(1/2) + \int_0^{\kappa x^2/D}U(1,3/2,z)dz + \mathcal{O}[s^2] \bigg{]}
\end{eqnarray} 
\end{small}
we get 

\begin{eqnarray}
\mathcal{T}_{\mathcal{HD}}(x_{th}|\underline{x_0,t_0})=\mathcal{T}_{\mathcal{HD}}(x_0)=\frac{1}{\kappa} \mathcal{T}^r_{\mathcal{HD}}(x_0)=\frac{1}{2\kappa}\int_{\kappa x_{th}^2 /D}^{\kappa x_{0}^2 /D} U(1,3/2,z)dz=\frac{\sqrt{\pi}}{\kappa} \int_{\sqrt{\kappa} |x_{th}|/\sqrt{D}}^{\sqrt{\kappa} |x_{0}|/\sqrt{D}} e^{y^2} \text{erfc}(y)dy\nonumber\\
=\frac{1}{{2 \kappa }} \bigg{[} {\pi  \text{erfi}\left(\frac{\sqrt{\kappa } x_0}{\sqrt{D}}\right)-\pi  \text{erfi}\left(\frac{\sqrt{\kappa } x_{\text{th}}}{\sqrt{D}}\right)} \bigg{]} -\frac{1}{{D}}\bigg{[}{x_0^2 \, _2F_2\left(1,1;\frac{3}{2},2;\frac{\kappa  x_0^2}{D}\right)-x_{\text{th}}^2 \, _2F_2\left(1,1;\frac{3}{2},2;\frac{\kappa  x_{\text{th}}^2}{D}\right)} \bigg{]} 
\end{eqnarray}
\end{widetext}

\section{Statistics of random excursions}
\label{appendix_exc}
\subsection{Statistics of sojourns above a threshold}
\label{peak-duration}
Let $x_{th}$ be the threshold of our interest and $x_{th}>x^*$. If $x$ denotes the current position of the Brownian particle, the time derivative $\dot{x}(t)$ denotes the rate of change of the position of the Brownian particle with time $t$. We now calculate the mean density $n(\tau^+)$ of sojourns whose duration exceed $\tau^+$. This will be followed by the evaluation of the corresponding probability distribution of sojourn times.

The problem is to count all such trajectories which upcross the threshold $x_{th}$ in the interval $[0,\Delta t]$ and do not downcross it in the next interval $[\Delta t, \tau^+]$. All such trajectories will give rise to sojourns whose duration exceed $\tau^+$. Let all such trajectories be described by the probability density $p^+(x,t)$. As these trajectories denote fluctuating position of the Brownian particle in time, probability density describing them will satisfy the following Fokker-Planck equation
\begin{eqnarray}
\frac{\partial p^+(x,t)}{\partial t}=\frac{\partial }{\partial x} \bigg{[} \kappa  x \ p^+(x,t)\bigg{]}+\frac{D}{2}\frac{\partial^2 p^+(x,t)}{\partial x^2}.  
\label{fpe-p-plus}
\end{eqnarray}
The above equation is subjected to the initial condition
\begin{equation}
p^+(x,0)=0~ \text{for} ~x>x_{th}
\label{ic_dur_exc}
\end{equation}
and boundary conditions
\begin{eqnarray}
p^+(x_{th},t)=\begin{cases}
& p_{ss}(x_{th}) ~ \text{for} ~0 \leq t \leq \Delta t \\
& 0 ~ \text{for} ~ \Delta t <t ~.
\label{bc_dur_exc}
\end{cases}
\end{eqnarray}
The initial condition (\ref{ic_dur_exc}) indicates that only those trajectories are considered   in which  the hypothetical Brownian particle is not above the threshold at $t=0$. The boundary condition (\ref{bc_dur_exc}) in the interval $[0,\Delta t]$ indicates that only those trajectories will contribute which are already at $x=x_{th}$ during the infinitesimal interval $[0,\Delta t]$. As the protrusion length is assumed to be in the stationary state, distribution of such trajectories are given by the stationary distribution $p_{ss}(x_{th})$ given in equation (\ref{pss-sol}). For $t > \Delta t$, the absorbing boundary condition $p^+(x_{th},t) = 0$ ensures elimination of all those trajectories whose sojourn time above $x_{th}$ would be less than $t$ because of premature downcrossing.

Integration of the probability $p^+(x, \tau^+)$, that satisfies the conditions (\ref{ic_dur_exc}) and (\ref{bc_dur_exc}), over the entire space above $x_{th}$ gives the number of sojourns $\Delta n$ above threshold which begin in $[0, \Delta t]$  and do not end by $t=\tau^+$, i.e., 
\begin{equation}
\Delta n= \int_{x_{th}}^ \infty p^+(x, \tau^+) dx. 
\end{equation}
Hence, the mean number density of sojourns above the threshold $x_{th}$, each of duration longer than $\tau^{+}$, is given by
\begin{equation}
n(\tau^+)=\lim_{\Delta t \to 0} \frac{\Delta n}{\Delta t}
\end{equation}
The  corresponding unnormalised probability density $p_n(\tau^+)$ of the duration of sojourn time above threshold is given by
\begin{equation}
p_n(\tau^+)=-\frac{dn(\tau^+)}{d \tau^+}
\label{eq-PnTau+}
\end{equation}
Solving the equation (\ref{fpe-p-plus}), subjected to the set of conditions given in equation (\ref{ic_dur_exc}) and (\ref{bc_dur_exc}), is a challenging task. Therefore, we will estimate the sojourn time distributions for rarely visited threshold $x_{th}$ ($x_{th}>>\sigma$) only.


For obtaining the expression for $n(\tau^+)$, we need the expression for $p^+(x,t)$ and the following limit
\begin{equation}
\lim_{t \to 0} \frac{p^+(x,t)}{\Delta t}
\label{lim1}
\end{equation}
For convenience , let us introduce the following function $w^+(x,t)$
\begin{equation}
w^+(x,t)=\frac{1}{p_{st}(x)}\lim_{t \to 0} \frac{p^+(x,t)}{\Delta t}
\label{lim2}
\end{equation}
Just as $p^+(x,t)$ satisfies the Fokker-Planck equation (\ref{fpe-p-plus}), this new function $w^+(x,t)$ also satisfies the following Fokker-Planck equation
\begin{eqnarray}
\frac{\partial w^+(x,t)}{\partial t}=\frac{\partial }{\partial x} \bigg{[} \kappa  x \ w^+(x,t)\bigg{]}+\frac{D}{2}\frac{\partial^2 w^+(x,t)}{\partial x^2}
\label{fpe_w_exc}  
\end{eqnarray}
and from the initial and boundary conditions (equation (\ref{ic_dur_exc}) and (\ref{bc_dur_exc}) respectively) we can infer that the equation (\ref{fpe_w_exc}) is subjected to the following conditions
\begin{eqnarray}
w^+(x,t)&=&0 ~ ~{\rm for}~ t<0, \nonumber \\
w^+(x_{th},t)&=&\delta(t).
\end{eqnarray}
In terms of $w^+(x,t)$, $n(\tau^+)$ is given by
\begin{equation}
n(\tau^+)=p_{st} \int_{x_{th}}^\infty w^+(x,\tau) dx 
\label{n_t_def}
\end{equation}
The partial differential equation (\ref{fpe_w_exc}) will be solved by using the Carlson-Laplace transform. The transform of $w^+(x,t)$ is given by
\begin{equation}
\tilde{w}^+(x,s)=s \int_0^\infty e^{-st} ~ {w}^+(x,t) dt 
\end{equation}

The Fokker-Planck equation (\ref{fpe_w_exc}) in terms of $\tilde{w}^+(x,s)$ is 
\begin{equation}
s \tilde{w}^+(x,s)=\frac{\partial}{\partial x} \bigg{[} \kappa x ~ \tilde{w}^+(x,s) \bigg{]}+\frac{D}{2}\frac{\partial^2 \tilde{w}^+(x,s)}{\partial x^2}    
\label{fpe_wt_exc}
\end{equation}
and the boundary condition takes the form
\begin{equation}
\tilde{w}^+(x_{th},s)=s~.
\label{bc_fpe_wt_exc}
\end{equation}
Using this boundary condition and the fact that the corresponding flux vanishes as $x\to \infty$, the Fokker-Planck equation (\ref{fpe_wt_exc}) can be integrated to
\begin{equation}
\int_{x_{th}}^\infty  \tilde{w}^+(x,s) dx =- \frac{1}{s} \bigg{[} \frac{D}{2}\frac{\partial \tilde{w}^+(x_{th},s)}{\partial x}+ \kappa x_{th} ~ \tilde{w}^+(x_{th},s)
\bigg{]}
\label{int_wt}
\end{equation}

Rearranging Eq.(\ref{fpe_wt_exc}),  we get
{\small 
\begin{equation}
 \frac{\partial^2 \tilde{w}^+(x,s)}{\partial x^2}+ \bigg{(}\frac{2\kappa}{D} \bigg{)} x  \frac{\partial \tilde{w}^+(x,s)}{\partial x}+\bigg{(}\frac{2\kappa}{D} \bigg{)}\bigg{(}1-\frac{s}{\kappa}\bigg{)}\tilde{w}^+(x,s)=0
\label{fpe_wt_exc_II}
\end{equation}}
which, upon change of variable from $x$ to $z$ defined by 
\begin{equation}
z=\sqrt{\frac{2\kappa}{D}} x,
\end{equation}
gets transformed to
\begin{equation}
 \frac{\partial^2 \tilde{w}^+(z,s)}{\partial z^2}+ z  \frac{\partial \tilde{w}^+(z,s)}{\partial z}+\bigg{(} 1 -\frac{s}{\kappa} \bigg{)}\tilde{w}^+(z,s)=0 ~.
\end{equation}
The general solution for the above equation has the form  
\begin{eqnarray}
\tilde{w}^+(x,s) = C_1 ~ e^{-z^2/4} \mathcal{D}_{-s/k}(z)+C_2 ~ e^{-z^2/4} \mathcal{D}_{-s/k}(-z) \nonumber \\
\end{eqnarray}
 where $\mathcal{D}_{-s/k}(y)$ is the parabolic cylindrical function and $C_1$ and $C_2$ are arbitrary constants fixed by the boundary conditions.
As the first term vanishes as $z \to \infty$  whereas the second does not, it qualifies as the  physically allowed solution. Imposing the boundary condition (\ref{bc_fpe_wt_exc}) at $x=x_{th}$ we get 
{\small \begin{equation}
\tilde{w}^+(x_{th},s)=s = C_1 ~ e^{-(\kappa x_{th}^2)/(2D)} \mathcal{D}_{-s/k}\biggl(\sqrt{\frac{2\kappa}{D}}x_{th}\biggr)
\label{bc_fpe_wt_exc5}
\end{equation}}
which gives the expression for $C_1$ that we use to write the solution
\begin{eqnarray}
\tilde{w}^+(x,s)=s  \exp \bigg{(} \frac{x_{th}^2 - x^2}{2D} \bigg{)} \frac{\mathcal{D}_{-s/k}(\sqrt{\frac{2\kappa}{D}} x)}{\mathcal{D}_{-s/k}(\sqrt{\frac{2\kappa}{D}} x_{th})}
\label{wt_expr_exc}
\end{eqnarray} 
The Carlson-Laplace transform of $n(\tau)$ is denoted by $\tilde{n}(s)$
and the expression of $\tilde{n}(s)$ using equation  (\ref{lim2}), (\ref{n_t_def}) and (\ref{int_wt}), we get 
{
\begin{eqnarray}
\tilde{n}(s)=  -  \frac{p_{st}(x)}{s} \bigg{[} \frac{D}{2} \frac{\partial \tilde{w}^+(x_{th},s)}{\partial x}+ \kappa x_{th} ~ \tilde{w}^+(x_{th},s)
\bigg{]} \nonumber\\
\label{n_t_expr}
\end{eqnarray} }

Substituting the expression for $\tilde{w}^+(x,s)$ in equation (\ref{n_t_expr}) and after some rearrangement we get
\begin{widetext}
{\small
\begin{eqnarray}
&\tilde{n}(x,s)& \nonumber\\ &=&-p_{st}(x_{th}) \nonumber\\ & \times & \bigg{[} \frac{D}{2\mathcal{D}_{-s/k} (\sqrt{\frac{2\kappa}{D})}x_{th})} ~ \exp\bigg{(} \frac{\kappa(x^2_{th}-x^2_{th})}{2D}\bigg{)} \bigg{(} \sqrt{\frac{2\kappa}{ D} } \bigg{)} \bigg{\{}- \frac{1}{2} \sqrt{\frac{2\kappa}{ D}}   {\mathcal{D}_{-s/k} \bigg{(}\sqrt{\frac{2\kappa}{D}}x_{th}\bigg{)}} +{\mathcal{D}'_{-s/k} \bigg{(}\sqrt{\frac{2\kappa}{D}}x_{th}\bigg{)}} \bigg{\}} + \kappa x_{th} \bigg{]} \nonumber\\ \nonumber\\
&(&\text{Using the identity $\mathcal{D}'_{\nu}(z)-\frac{z}{2} \mathcal{D}_\nu (z)+ \mathcal{D}_{\nu +1}(z)=0$}~)\nonumber\\ \nonumber\\
&=&-p_{st}(x_{th}) \bigg{[} \frac{D}{2\mathcal{D}_{-s/k} (\sqrt{\frac{2\kappa}{D})}x_{th})} ~  \bigg{(} \sqrt{\frac{2\kappa}{ D} } \bigg{)} \bigg{\{}   {-\mathcal{D}_{-s/k+1} \bigg{(}\sqrt{\frac{2\kappa}{D}}x_{th}\bigg{)}}  \bigg{\}} + \kappa x_{th} \bigg{]} \nonumber\\ \nonumber\\
&(&\text{Using the identity $\mathcal{D}_{\nu+1}(z)-z \mathcal{D}_\nu (z)+ \nu \mathcal{D}_{\nu -1}(z)=0$} ~~)\nonumber\\ \nonumber\\
&=&-p_{st}(x_{th}) \bigg{[} \frac{D}{2\mathcal{D}_{-s/k} (\sqrt{\frac{2\kappa}{D})}x_{th})} ~  \bigg{(}  \sqrt{\frac{2\kappa}{ D} } \bigg{)} ~  \bigg{\{}   {-\bigg{(}\sqrt{\frac{2\kappa}{D}}x_{th} \bigg{)} \mathcal{D}_{-s/k} \bigg{(}\sqrt{\frac{2\kappa}{D}}x_{th}\bigg{)} + (-s/k) \mathcal{D}_{-s/k-1} \bigg{(}\sqrt{\frac{2\kappa}{D}}x_{th}\bigg{)}}  \bigg{\}} + \kappa x_{th} \bigg{]} \nonumber\\ \nonumber\\
&=& -p_{st}(x_{th}) \bigg{[} -s ~\sqrt{\frac{D}{2\kappa}} ~ \frac{\mathcal{D}_{-s/k-1} \bigg{(}\sqrt{\frac{2\kappa}{D}}x_{th}\bigg{)}}{\mathcal{D}_{-s/k} \bigg{(}\sqrt{\frac{2\kappa}{D}}x_{th}\bigg{)}}  \bigg{]} =  s ~\sqrt{\frac{D}{2\kappa}} ~ p_{st}(x_{th})  ~ \frac{\mathcal{D}_{-s/k-1} \bigg{(}\sqrt{\frac{2\kappa}{D}}x_{th}\bigg{)}}{\mathcal{D}_{-s/k} \bigg{(}\sqrt{\frac{2\kappa}{D}}x_{th}\bigg{)}} 
\label{nt_expr_exc_II}
\end{eqnarray} }
\end{widetext}

In case of higher thresholds ($x_{th}>>D/2\kappa$), the drift that acts on the Brownian particle which makes shorter excursion beyond the threshold $x_{th}$ can be approximated as a $\kappa x \approx \kappa x_{th}$. So the equation (\ref{fpe_wt_exc_II}) simplifies to
\begin{equation}
s \tilde{w}^+(x,s)=\kappa x_{th} \frac{\partial \tilde{w}^+(x,s)}{\partial x} +\frac{D}{2}\frac{\partial^2 \tilde{w}^+(x,s)}{\partial x^2}  
\end{equation}
whose general solution is
\begin{eqnarray}
&\tilde{w}&^+(x,s)=C_1 \exp \bigg{( } -\frac{x}{D}(\kappa x_{th} + \sqrt{(\kappa x_{th})^2+2sD}) \bigg{) } \nonumber\\ &+& C_2 \exp  \bigg{( } -\frac{x}{D}(\kappa x_{th} - \sqrt{(\kappa x_{th})^2+2sD})) \bigg{)} 
\end{eqnarray}
Using the boundary condition (equation \ref{bc_fpe_wt_exc}), we get
\begin{eqnarray}
\tilde{w}^+(x,s)=s \times \nonumber\\   \exp \bigg{(} \frac{(x_{th}-x)}{D} (\kappa x_{th} - \sqrt{(\kappa x_{th})^2+2sD}) \bigg{)}
\end{eqnarray}
Using it in equation (\ref{n_t_expr}), we get 
\begin{equation}
\tilde{n}(s)=p_{st}(x) \frac{s D}{\kappa x_{th} + \sqrt{(\kappa x_{th})^2 + 2sD}}
\end{equation}
The inverse Carlson-Laplace transform is given by
\begin{eqnarray}
n(\tau)&=&\frac{p_{st}(x_{th})}{2}  \bigg{[ } \frac{1}{{\sqrt{\pi \tau /2D}}}~{e^{-(\kappa x_{th})^2 \tau/(2D)}}  \nonumber\\ &-& (\kappa x_{th}) ~ \text{erfc} \bigg{(} \kappa x_{th} \sqrt{\frac{\tau}{2D}} \bigg{)} \bigg{]}
\end{eqnarray}

\section{Extreme excursions}
\label{extreme-exc-app}

The probability of $x_{max}(t)$ being the maximum is denoted by 
\begin{equation}
\mathcal{M}_{max}(t;\zeta|\underline{x_0,t_0})=\text{Probability}[x_{max}(t)<\zeta|\underline{x_0,t_0}]
\label{max-def}
\end{equation}
and the  probability of $x_{min}(t)$ being the minimum is denoted by 
\begin{equation}
\mathcal{M}_{min}(t;\zeta|\underline{x_0,t_0})=\text{Probability}[\zeta<x_{min}(t)|\underline{x_0,t_0}]~.
\label{min-def}
\end{equation}
For simplicity, we will keep our discussion limited to the calculation of statistics of $x_{max}$. Once we get it for the maximum, getting the results for the minimum $x_{min}$ will be straight forward. 

In equation (\ref{max-def}), the term $\text{Probability}[x_{max}(t)<\zeta|\underline{x_0,t_0}]$ means that the particle which was initially at $x_0$ ($-\infty < x_0 < \zeta$) has not escaped the region $[-\infty,\zeta]$. In other words, it has not hit the threshold $x_{th}=\zeta$  till time $t$. Therefore,
\begin{equation}
\mathcal{M}_{max}(t;\zeta|\underline{x_0,t_0})=(1-\mathcal{H}_U(t;\zeta| \underline{x_0,t_0}))\Theta(x_0-\zeta)~.
\label{max-def-alt}
\end{equation}
where $\mathcal{H}_U(t;\zeta| \underline{x_0,t_0})$ is the upcrossing probability for the threshold $\zeta$ which was introduced in equation (\ref{upc-prob}). The Heaviside function $\Theta(x_0-\zeta)$ takes care of the fact that if initially the particle is located beyond $\zeta$, then it is impossible to have $\zeta$ as the maximum. As the {\it pdf} corresponding to $\mathcal{M}_{max}(t;\zeta|\underline{x_0,t_0})$ is $p_{max}(t;\zeta|\underline{x_0,t_0})$ , the mean maximum $\langle x_{max}(t) | x_0\rangle$ is given by
\begin{equation}
\langle x_{max}(t) | x_0\rangle = \int_0 ^\infty \zeta ~ p_{max}(t;\zeta|\underline{x_0,t_0})~ d\zeta~.
\end{equation}

Rather than having arbitrary values of $x_0$, we fix $x_0=0(=x^*)$  without loss of generality. This will give us the average maximum excursion $\langle x_{max}(t) | x_0\rangle $ about the mean position $x^*=0$ in steady state.

The {\it pdf} $p_{max}(t;\zeta|\underline{x_0,t_0})$ is given by
\begin{eqnarray}
p_{max}(t;\zeta|\underline{x_0,t_0})=\frac{\partial  \mathcal{M}_{max}(t;\zeta|\underline{x_0,t_0}) }{\partial  \zeta} \nonumber\\=-\frac{\partial  \mathcal{H}_U(t;\zeta|\underline{x_0,t_0}) }{\partial  \zeta} \Theta(\zeta-x_0)
\end{eqnarray}
 where we have used (\ref{max-def-alt}) in the second step.
Therefore,
\begin{eqnarray}
\langle {x}_{max} (t) |x_0 \rangle= \int_{-\infty} ^\infty \zeta ~ p_{max}(t;\zeta|\underline{x_0,t_0})~ d\zeta~ \nonumber\\ =x_0 + \int_{x_0}^\infty \mathcal{H}_U(t;\zeta|\underline{x_0,t_0})d \zeta \nonumber\\
\end{eqnarray}
For $x_0=0$, from the above equation we get
\begin{eqnarray}
\langle {x}_{max} (t) | 0 \rangle =  \int_{0}^\infty \mathcal{H}_U(t;\zeta|\underline{0,t_0})d \zeta ~.
\end{eqnarray}
In the above equation, replacing the time variable $t$ with the rescaled time variable  $t_r$ where $t_r=\kappa t$, we get
\begin{eqnarray}
\langle {x}_{max} (t_r) | 0 \rangle =  \int_{0}^\infty \mathcal{H}_U(t_r;\zeta|\underline{0,t_{r0}})d \zeta 
\end{eqnarray}

Taking the the Laplace transform of both the sides
\begin{eqnarray}
\langle {\tilde{x}}_{max} (s) | 0 \rangle =  \int_{0}^\infty \mathcal{\tilde{H}}_U(s;\zeta|\underline{0})~d \zeta \nonumber\\
=\frac{1}{s} \int_{0}^\infty   \frac{F(s/2,1/2,0)}{ F(s/2,1/2, \kappa \zeta^2/D )} ~d \zeta
\label{ext-exc-integral}
\end{eqnarray}
where we have used the expression of $\mathcal{\tilde{H}}_U(s;\zeta|\underline{0})$ which was already  derived  in appendix \ref{app-mht} and given in equation (\ref{hp_sol}). Using the following value,
\begin{equation}
F(s/2,1/2,0)=0
\end{equation}
and using the following expansion
\begin{equation}
{ F(s/2,1/2, \kappa \zeta^2/D )}=1+s \frac{\kappa \zeta^2 }{D} 
\end{equation}
for small $\zeta$
the integral in equation (\ref{ext-exc-integral}) simplifies to
\begin{equation}
\langle {\tilde{x}}_{max} (s) | 0 \rangle = \frac{1}{s} \int_{0}^\infty   \frac{d \zeta}{1+(s {\kappa \zeta^2 /D})}=\frac{\pi}{2} \sqrt{\frac{D}{\kappa}} s^{-3/2}
\end{equation}
Carrying out the inverse Laplace transform we get 

\begin{equation}
\langle {{x}}_{max} (t_r) | 0 \rangle = \left( \pi {\frac{D}{\kappa}} {t_r} \right)^{1/2}.
\end{equation}
Finally, restoring physical time $t$, i.e., replacing $t_r$ by $\kappa t$, we get
\begin{equation}
\langle {{x}}_{max} (t) | 0 \rangle = \sqrt{ \pi D t }~.
\end{equation}

As derived in detail in appendix \ref{extreme-exc-app}, it is shown that the average maximum $\langle {x}_{max} (t) |0 \rangle $  goes as
\begin{equation}
\langle {x}_{max} (t) |0 \rangle \simeq \sqrt{{\pi} Dt}
\label{xmax}
\end{equation}
i.e, the $\langle {x}_{max} (t) |0 \rangle \propto \sqrt{Dt}$. 

Next, let us calculate $\langle {x}_{min} (t) |0 \rangle$. If initially $x_0=0$, then $\langle x_{min}(t)|0 \rangle$ is related to the first downcrossing time to $x_{th}=-\zeta$.   We also know that 
\begin{equation}
\mathcal{H}_D(t;-\zeta| \underline{0,t_0})=\mathcal{H}_U(t;\zeta| \underline{0,t_0}).
\end{equation}
Therefore, without going through detailed calculations again, we can simply write
\begin{equation}
\langle {x}_{min} (t) |0 \rangle=-\langle {x}_{max} (t) |0 \rangle=- \sqrt{\pi D t} ~ .
\label{xmin}
\end{equation}

The range scanned by the Brownian particle is also a time dependent random variable  that depends on both $x_{max}(t)$ and $x_{min}(t)$. The average  width of the range is denoted by $\langle x_{range}(t)|0 \rangle$. It can be evaluated directly by subtracting the average minimum from the average maximum i.e,
\begin{equation}
\langle x_{range}(t)|0 \rangle=\langle {x}_{max} (t) |0 \rangle -\langle {x}_{min} (t) |0 \rangle =2\sqrt{ \pi Dt }
\label{xrange}
\end{equation}



\begin{thebibliography}{99} 

\bibitem{masoliverbook} J. Masoliver, {\it Random processes: first passage and escape} (World Scientific, 2018).

\bibitem{rednerbook} S. Redner {\it A guide to first-passage processes} (Cambridge University Press, 2001). 

\bibitem{masoliver14} Masoliver, Jaume. “The Level-Crossing Problem: First-Passage, Escape and Extremes.” {\it Fluctuation and Noise Letters}, vol. 13, no. 04, 2014, p. 1430001., doi:10.1142/

\bibitem{braininabook}Brainina, Irina S. Applications of Random Process Excursion Analysis. Elsevier, 2017.



\bibitem{nordin70} C.F. Nordin and D.M. Rosberg, {\it Applications of crossing theory in hydrology}, Hydrological Sciences Journal, {\bf 15}, 27-43 (1970). 



\bibitem{ghusinga17}Ghusinga, Khem Raj et al. “First-passage time approach to controlling noise in the timing of intracellular events.” Proceedings of the National Academy of Sciences of the United States of America vol. 114,4 (2017): 693-698. doi:10.1073/pnas.1609012114

\bibitem{thorneywork19}Thorneywork, Alice L., et al. “Direct Detection of Molecular Intermediates from First-Passage Times.” 2019, doi:10.1101/772830.

\bibitem{dhar19}Dhar, Abhishek, et al. “Run-and-Tumble Particle in One-Dimensional Confining Potentials: Steady-State, Relaxation, and First-Passage Properties.” Physical Review E, vol. 99, no. 3, 2019, doi:10.1103/physreve.99.032132.

\bibitem{besga20}Besga, Benjamin, et al. “Optimal Mean First-Passage Time for a Brownian Searcher Subjected to Resetting: Experimental and Theoretical Results.” Physical Review Research, vol. 2, no. 3, 2020, doi:10.1103/physrevresearch.2.032029.

\bibitem{ghosh18}Ghosh, Soumendu, et al. “First-Passage Processes on a Filamentous Track in a Dense Traffic: Optimizing Diffusive Search for a Target in Crowding Conditions.” Journal of Statistical Mechanics: Theory and Experiment, vol. 2018, no. 12, 2018, p. 123209., doi:10.1088/1742-5468/aaf31d.

\bibitem{bel20}Bel, G., Zilman, A. \&  Kolomeisky, A.B. “Different time scales in dynamic systems with multiple exits.” (2020) arXiv:2006.10613  

\bibitem{evans14}Evans, Martin R, and Satya N Majumdar. “Diffusion with Resetting in Arbitrary Spatial Dimension.” Journal of Physics A: Mathematical and Theoretical, vol. 47, no. 28, 2014, p. 285001., doi:10.1088/1751-8113/47/28/285001.

\bibitem{zhang16} Zhang, Yaojun, and Olga K Dudko. “First-Passage Processes in the Genome.” Annual review of biophysics vol. 45 (2016): 117-34. doi:10.1146/annurev-biophys-062215-010925

\bibitem{polizzi16}Polizzi, Nicholas F et al. “Mean First-Passage Times in Biology.” Israel journal of chemistry vol. 56,9-10 (2016): 816-824. doi:10.1002/ijch.201600040

\bibitem{metzler14} Metzler, R., Oshanin, G., \&; Redner, S. (2014). First-passage phenomena and their applications. New Jersey: World Scientific.

\bibitem{iyerbiswas16} Iyer‐Biswas, S. and Zilman, A. (2016). First‐Passage Processes in Cellular Biology. In Advances in Chemical Physics (eds S.A. Rice and A.R. Dinner). doi:10.1002/9781119165156.ch5

\bibitem{guillet19} Guillet, A., Roldan, E.,  Julicher , F.,  Extreme-Value Statistics of Molecular Motors arXiv preprint arXiv:1908.03499 (2019)
\bibitem{greulich18}Greulich, P., \&; Simons, B. D. (2018). Extreme value statistics of mutation accumulation in renewing cell populations. Physical Review E, 98(5). doi:10.1103/physreve.98.050401

\bibitem{syski92} Syski, Ryszard. {\it Passage Times for Markov Chains}. IOS Press, 1992.

\bibitem{stratonovic81} Stratonovic R. L.  {\it Topics in the Theory of Random Noise.} Gordon and Breach, 1981.

\bibitem{malakar18} Malakar, Kanaya, et al. “Steady State, Relaxation and First-Passage Properties of a Run-and-Tumble Particle in One-Dimension.” Journal of Statistical Mechanics: Theory and Experiment, vol. 2018, no. 4, 2018, p. 043215., doi:10.1088/1742-5468/aab84f.

\bibitem{zacks17}Zacks, Shelemyahu. Sample Path Analysis and Distributions of Boundary Crossing Times. Springer International Publishing, 2017.


\bibitem{patra20} Patra, Swayamshree, et al. “Flagellar Length Control in Biflagellate Eukaryotes: Time-of-Flight, Shared Pool, Train Traffic and Cooperative Phenomena.” New Journal of Physics, vol. 22, no. 8, (2020), p. 083009., doi:10.1088/1367-2630/ab9ee4.


\bibitem{mohapatra16}Mohapatra, Lishibanya et al. “Design Principles of Length Control of Cytoskeletal Structures.” Annual review of biophysics vol. 45 (2016): 85-116. doi:10.1146/annurev-biophys-070915-094206


\bibitem{marshall01} Marshall, W F, and J L Rosenbaum. "Intraflagellar transport balances continuous turnover of outer doublet microtubules: implications for flagellar length control.” The Journal of cell biology vol. 155,3 (2001): 405-14. doi:10.1083/jcb.200106141

\bibitem{chu17}Chu, Fang-Yi et al. “On the origin of shape fluctuations of the cell nucleus." Proceedings of the National Academy of Sciences of the United States of America vol. 114,39 (2017): 10338-10343. doi:10.1073/pnas.1702226114

\bibitem{amiri20} Amiri, K. P. et al. Robustness and universality in organelle size control. bioRxiv 789453; doi: https://doi.org/10.1101/789453


\bibitem{mukherji14}Mukherji, Shankar, and Erin K O'Shea. “Mechanisms of organelle biogenesis govern stochastic fluctuations in organelle abundance.” eLife vol. 3 e02678. 10 Jun. 2014, doi:10.7554/eLife.02678


\bibitem{yuan17}Yuan, Aidong et al. "Neurofilaments and Neurofilament Proteins in Health and Disease.” Cold Spring Harbor perspectives in biology vol. 9,4 a018309. 3 Apr. 2017, doi:10.1101/cshperspect.a018309

\bibitem{herrmann16}Herrmann, Harald, and Ueli Aebi. "Intermediate Filaments: Structure and Assembly.” Cold Spring Harbor perspectives in biology vol. 8,11 a018242. 1 Nov. 2016, doi:10.1101/cshperspect.a018242

\bibitem{pollard03}Pollard TD, Borisy GG. Cellular motility driven by assembly and disassembly of actin filaments [published correction appears in Cell. 2003 May 16;113(4):549]. Cell. 2003;112(4):453-465. doi:10.1016/s0092-8674(03)00120-x

\bibitem{borisy16}Borisy, Gary et al. "Microtubules: 50 years on from the discovery of tubulin." Nature reviews. Molecular cell biology vol. 17,5 (2016): 322-8. doi:10.1038/nrm.2016.45

\bibitem{piao09}Piao, Tian et al. "A microtubule depolymerizing kinesin functions during both flagellar disassembly and flagellar assembly in Chlamydomonas.” PNAS vol. 106,12 (2009): 4713-8. doi:10.1073/pnas.0808671106



\bibitem{haldane25} J.B.S. Haldane, {\it On being the right size}, Harper's magazine (March, 1926). 



\bibitem{bonnerbook} J. T. Bonner, {\it Why Size matters: From Bacteria to Blue Whales} (Princeton University Press, 2006).

\bibitem{hamant20} Hamant O, Saunders TE. Shaping Organs: Shared Structural Principles Across Kingdoms [published online ahead of print, 2020 Jul 6]. Annu Rev Cell Dev Biol. 2020;10.1146/annurev-cellbio-012820-103850. doi:10.1146/annurev-cellbio-012820-103850

\bibitem{marshall15a}Marshall, Wallace F. “How Cells Measure Length on Subcellular Scales.” Trends in cell biology vol. 25,12 (2015): 760-768. doi:10.1016/j.tcb.2015.08.008

\bibitem{marshall15b}Marshall, Wallace F. "Subcellular size.” Cold Spring Harbor perspectives in biology vol. 7,6 a019059. 8 May. 2015, doi:10.1101/cshperspect.a019059

\bibitem{marshall16}Marshall, Wallace F. "Cell Geometry: How Cells Count and Measure Size.” Annual review of biophysics vol. 45 (2016): 49-64. doi:10.1146/annurev-biophys-062215-010905

\bibitem{rafelski08}Rafelski, Susanne M, and Wallace F Marshall. “Building the cell: design principles of cellular architecture.” Nature reviews. Molecular cell biology vol. 9,8 (2008): 593-602. doi:10.1038/nrm2460

\bibitem{ludington15} Ludington, William B et al. “A systematic comparison of mathematical models for inherent measurement of ciliary length: how a cell can measure length and volume.” Biophysical journal vol. 108,6 (2015): 1361-1379. doi:10.1016/j.bpj.2014.12.051


\bibitem{albus13}Albus, Christin A et al. “Cell length sensing for neuronal growth control.” Trends in cell biology vol. 23,7 (2013): 305-10. doi:10.1016/j.tcb.2013.02.001

\bibitem{folz19}Folz, Frederic et al. “Sound of an axon's growth.” Physical review. E vol. 99,5-1 (2019): 050401. doi:10.1103/PhysRevE.99.050401


\bibitem{wordeman09} Wordeman, Linda, and Jason Stumpff. “Microtubule length control, a team sport?.” Developmental cell vol. 17,4 (2009): 437-8. doi:10.1016/j.devcel.2009.10.002

\bibitem{melbinger12} Melbinger, Anna et al. “Microtubule length regulation by molecular motors.” Physical review letters vol. 108,25 (2012): 258104. doi:10.1103/PhysRevLett.108.258104

\bibitem{rank18} Rank, Matthias et al. Limited Resources Induce Bistability in Microtubule Length Regulation.” Physical review letters vol. 120,14 (2018): 148101. doi:10.1103/PhysRevLett.120.148101

\bibitem{johann12}Johann, Denis et al. Length regulation of active biopolymers by molecular motors.” Physical review letters vol. 108,25 (2012): 258103. doi:10.1103/PhysRevLett.108.258103

\bibitem{klein05} Klein, Gernot A et al. “Filament depolymerization by motor molecules.” Physical review letters vol. 94,10 (2005): 108102. doi:10.1103/PhysRevLett.94.108102

\bibitem{govindan08} Govindan, B. S., et al. “Length Control of Microtubules by Depolymerizing Motor Proteins.” EPL (Europhysics Letters), vol. 83, no. 4, 2008, p. 40006., doi:10.1209/0295-5075/83/40006.

\bibitem{kuan13} Kuan, Hui-Shun, and M D Betterton. "Biophysics of filament length regulation by molecular motors.” Physical biology vol. 10,3 (2013): 036004. doi:10.1088/1478-3975/10/3/036004


\bibitem{mohapatra15} Mohapatra, Lishibanya et al. "Antenna Mechanism of Length Control of Actin Cables.” PLoS computational biology vol. 11,6 e1004160. 24 Jun. 2015, doi:10.1371/journal.pcbi.1004160

\bibitem{varga09}Varga, Vladimir et al. Kinesin-8 motors act cooperatively to mediate length-dependent microtubule depolymerization.” Cell vol. 138,6 (2009): 1174-83. doi:10.1016/j.cell.2009.07.032 


\bibitem{orly15} Orly, Gilad et al. “A Biophysical Model for the Staircase Geometry of Stereocilia.  PloS one vol. 10,7 e0127926. 24 (2015), doi:10.1371/journal.pone.0127926


\bibitem{prost07}Prost, Jacques et al. “Dynamical control of the shape and size of stereocilia and microvilli.” Biophysical journal vol. 93,4 (2007): 1124-33. doi:10.1529/biophysj.106.098038

\bibitem{reese14} Reese, Louis et al. “Molecular mechanisms for microtubule length regulation by kinesin-8 and XMAP215 proteins.” Interface focus vol. 4,6 (2014): 20140031. doi:10.1098/rsfs.2014.0031

\bibitem{erlenkamper09} Erlenkämper, C, and K Kruse. “Uncorrelated Changes of Subunit Stability Can Generate Length-Dependent Disassembly of Treadmilling Filaments.” Physical Biology, vol. 6, no. 4, 2009, p. 046016., doi:10.1088/1478-3975/6/4/046016.

\bibitem{fai19} Fai, Thomas G et al. “Length regulation of multiple flagella that self-assemble from a shared pool of components.” eLife vol. 8 e42599. 9 Oct. 2019, doi:10.7554/eLife.42599

\bibitem{gov06} Gov, Nir S. “Dynamics and morphology of microvilli driven by actin polymerization." Physical review letters vol. 97,1 (2006): 018101. doi:10.1103/PhysRevLett.97.018101


\bibitem{banerjee20} Banerjee, Deb Sankar, and Shiladitya Banerjee. “Size Regulation of Multiple Organelles Competing for a Shared Subunit Pool.” 2020, doi:10.1101/2020.01.11.902783.


\bibitem{ma20}Ma, Rui et al. “Speed and Diffusion of Kinesin-2 Are Competing Limiting Factors in Flagellar Length-Control Model.” Biophysical journal vol. 118,11 (2020): 2790-2800. doi:10.1016/j.bpj.2020.03.034


\bibitem{bauer20} Bauer, D., Ishikawa, H., Wemmer, K. A., Hendel, N. L., Kondev, J., \& Marshall, W. F. (2021). Analysis of biological noise in the flagellar length control system. iScience, 24(4), 102354. https://doi.org/10.1016/j.isci.2021.102354


\bibitem{mogilner96} Mogilner, A, and G Oster. “Cell motility driven by actin polymerization.” Biophysical journal vol. 71,6 (1996): 3030-45. doi:10.1016/S0006-3495(96)79496-1

\bibitem{mogilner03}Mogilner A, Oster G. Polymer motors: pushing out the front and pulling up the back. Curr Biol. 2003;13(18):R721-R733. doi:10.1016/j.cub.2003.08.050

\bibitem{mcgrath17}McGrath, Jamis et al. “Stereocilia morphogenesis and maintenance through regulation of actin stability.” Seminars in cell \& developmental biology vol. 65 (2017): 88-95. doi:10.1016/j.semcdb.2016.08.017

\bibitem{ortega19}Vélez-Ortega, A Catalina, and Gregory I Frolenkov. “Building and repairing the stereocilia cytoskeleton in mammalian auditory hair cells.” Hearing research vol. 376 (2019): 47-57. doi:10.1016/j.heares.2018.12.012

\bibitem{brown10}Brown, Jeffrey W, and C James McKnight. “Molecular model of the microvillar cytoskeleton and organization of the brush border.” PloS one vol. 5,2 e9406. 24 Feb. 2010, doi:10.1371/journal.pone.0009406

\bibitem{snell04}Snell, William J et al. “Cilia and flagella revealed: from flagellar assembly in Chlamydomonas to human obesity disorders.” Cell vol. 117,6 (2004): 693-7. doi:10.1016/j.cell.2004.05.019

\bibitem{ferreira19} R Ferreira, R., Fukui, H., Chow, R., Vilfan, A., \& Vermot, J. (2019). The cilium as a force sensor-myth versus reality. Journal of cell science, 132(14), jcs213496. https://doi.org/10.1242/jcs.213496

\bibitem{kratz06}Kratz, Marie F. “Level Crossings and Other Level Functionals of Stationary Gaussian Processes.” Probability Surveys, vol. 3, 2006, pp. 230–288., doi:10.1214/154957806000000087.

\bibitem{rice44} Rice, S. O. “Mathematical Analysis of Random Noise.” Bell System Technical Journal, vol. 23, no. 3, 1944, pp. 282–332., doi:10.1002/j.1538-7305.1944.tb00874.x.

\bibitem{rice45} Rice, S. O. “Mathematical Analysis of Random Noise.” Bell System Technical Journal, vol. 24, no. 1, 1945, pp. 46–156., doi:10.1002/j.1538-7305.1945.tb00453.x.

\bibitem{khona13}Khona, D. K., Rao, V. G., Motiwalla, M. J., Varma, P. C., Kashyap, A. R., Das, K., Shirolikar, S. M., Borde, L., Dharmadhikari, J. A., Dharmadhikari, A. K., Mukhopadhyay, S., Mathur, D., \& D'Souza, J. S. (2013). Anomalies in the motion dynamics of long-flagella mutants of Chlamydomonas reinhardtii. Journal of biological physics, 39(1), 1–14. https://doi.org/10.1007/s10867-012-9282-8

\bibitem{milo16} Milo, Ron, et al. Cell Biology by the Numbers. Garland Science, 2016.


\bibitem{ishikawa17} H. Ishikawa and W.F. Marshall, { Testing the time-of-flight model for flagellar length sensing}, 
Mol. Biol. Cell {\bf 28}, 3447-3456 (2017).


\bibitem{lechtreck17} K.F. Lechtrack, J.C. Van De Weghe, J.A. Harris and P. Liu, {\it Protein transport in growing and steady-state cilia}, Traffic {\bf 18}, 277-286 (2017).


\bibitem{engel09} B.D. Engel, W.B. Ludington and W.F. Marshall, {\it Intraflagellar transport particle size scales inversely with flagellar length: revisiting the balance-point length control model}, J. Cell Biol. {\bf 187}, 81-89 (2009). 

\bibitem{kozminski93} K.G. Kozminski,  K.A. Johnson, P. Forscher and J.L. Rosenbaum, {\it A motility in the eukaryotic flagellum unrelated to flagellar beating}, PNAS {\bf 90}, 5519-5523 (1993). 

\bibitem{kozminski12} K.G. Kozminski, {\it Intraflagellar transport- the ``new motility'' 20 years later}, Mol. Biol. Cell {\bf 23}, 751-753 (2012). 

\bibitem{rosenbaum02} J.L. Rosenbaum and G.B. Witman, {\it Intraflagellar transport}, Nat. Rev. Mol. Cell Biol. {\bf 3}, 813 (2002). 



\bibitem{stepanek16} L. Stepanek and G. Pigino, {\it Microtubule doublets are double-track railways for Intraflagellar transport trains }, Science.  {\bf 352}, 721-4 (2016).

\bibitem{wren13} Wren, K. N., Craft, J. M., Tritschler, D., Schauer, A., Patel, D. K., Smith, E. F., Porter, M. E., Kner, P., \& Lechtreck, K. F. (2013). A differential cargo-loading model of ciliary length regulation by IFT. Current biology : CB, 23(24), 2463–2471. https://doi.org/10.1016/j.cub.2013.10.044

\bibitem{ludington13} Ludington, W. B., Wemmer, K. A., Lechtreck, K. F., Witman, G. B., \& Marshall, W. F. (2013). Avalanche-like behavior in ciliary import. Proceedings of the National Academy of Sciences of the United States of America, 110(10), 3925–3930. https://doi.org/10.1073/pnas.1217354110


\bibitem{bressloff18} Bressloff P. C. \& Karamched. B. R. (2018).Doubly Stochastic Poisson Model of Flagellar Length Control. SIAM J. Appl. Math., 78(2), 719–741

\bibitem{bressloff06} Bressloff P. C. (2006). Stochastic model of intraflagellar transport. Physical review. E, Statistical, nonlinear, and soft matter physics, 73(6 Pt 1), 061916. https://doi.org/10.1103/PhysRevE.73.061916

\bibitem{patra18} Patra, S., \& Chowdhury, D. (2018). Multispecies exclusion process with fusion and fission of rods: A model inspired by intraflagellar transport. Physical review. E, 97(1-1), 012138. https://doi.org/10.1103/PhysRevE.97.012138
 


\bibitem{kampen10}Kampen, N. G. van. Stochastic Processes in Physics and Chemistry. World Publishing Corporation, 2010.

\bibitem{gardiner09}Gardiner, Crispin W. Stochastic Methods: a Handbook for the Natural and Social Sciences. Springer, 2009.

\bibitem{gillespie13}Gillespie, Daniel T., and Effrosyni Seitaridou. Simple Brownian Diffusion: an Introduction to the Standard Theoretical Models. Oxford University Press, 2013.

\bibitem{abramowitz72} Abramowitz, Milton, and Irene A. Stegun. Handbook of Mathematical Functions with Formulas, Graphs, and Mathematical Tables. U.S. Dept. of Commerce, National Bureau of Standards, 1972.

\bibitem{hartich19} Hartich, David, and Aljaž Godec. “Extreme Value Statistics of Ergodic Markov Processes from First Passage Times in the Large Deviation Limit.” Journal of Physics A: Mathematical and Theoretical, vol. 52, no. 24, 2019, p. 244001., doi:10.1088/1751-8121/ab1eca.


\bibitem{besschetnova10} Besschetnova, T. Y., Kolpakova-Hart, E., Guan, Y., Zhou, J., Olsen, B. R., \& Shah, J. V. (2010). Identification of signaling pathways regulating primary cilium length and flow-mediated adaptation. Current biology : CB, 20(2), 182-187. https://doi.org/10.1016/j.cub.2009.11.072

\bibitem{hendel18} Hendel, N. L., Thomson, M., \& Marshall, W. F. (2018). Diffusion as a Ruler: Modeling Kinesin Diffusion as a Length Sensor for Intraflagellar Transport. Biophysical journal, 114(3), 663–674. https://doi.org/10.1016/j.bpj.2017.11.3784


‌‌\bibitem{mcinally19} Mcinally, Shane G, et al. ``Length Dependent Disassembly Maintains Four Different Flagellar Lengths in Giardia.'' \textit{ ELife}, vol. 8, 2019, doi:10.7554/elife.48694.

\end{thebibliography}
\end{document}